\documentclass[10pt,a4paper]{article}
\usepackage[utf8]{inputenc}
\usepackage{amsmath}
\usepackage{amsfonts}
\usepackage{amssymb}
\usepackage{graphicx}

\usepackage{a4wide}

\newtheorem{theorem}{Theorem}
\newtheorem{statement}{Statement}
\newtheorem{lemma}{Lemma}

\pdfoutput=1

\title{Asymptotics of the whispering gallery-type in the eigenproblem for the Laplacian in a revolutional domain diffeomorphic to a solid torus}

\author{D.S. Minenkov\footnote{Mechanics and Mathematics Faculty of M.V. Lomonosov Moscow State University, Moscow, Russia;
Ishlinsky Institute for Problems in Mechanics RAS, Moscow, Russia, minenkov.ds@gmail.com }, S.A. Sergeev\footnote{Pontifical Catholic University of Rio de Janeiro, PUC-Rio, Rio de Janeiro, Brazil, sergeevse1@gmail.com}}

\date{}

\begin{document}
\maketitle
\begin{abstract}\noindent
We consider the eigenproblem for the Laplacian inside a three-dimensional revolutional domain diffeomorphic to a solid torus and construct asymptotic eigenvalues and eigenfunctions (quasimodes) of the whispering gallery-type. The whispering gallery-type asymptotics are localized near the boundary of
the domain, and an explicit analytic representations in terms of Airy
functions is constructed for such asymptotics. There are several different scales in the problem, which makes it possible to apply the procedure of adiabatic approximation in the form of operator separation of variables to reduce the initial problem to one-dimensional problems up to the small correction. We also discuss the relation between the constructed whispering gallery-type asymptotics and classical billiards in the corresponding domain, in particularly,  such asymptotics correspond to almost integrable billiards with proper degeneracy. We illustrate the results in the case when a revolutional domain is obtained by rotation of the triangle with rounded wedges.

{\bf Keywords:} Laplace operator, spectral problem, solid torus, quasimodes, localized eigenfunctions, whispering gallery, adiabatic approximation, operator separation of variables, Maslov canonical operator, semiclassical asymptotics, Airy function

{\bf MSC 2020:} 35J05, 35B40, 35C20, 35P20, 81Q20

\end{abstract}

\section{Introduction.}
The effect of whispering galleries in acoustics has been known for a long time period and was studied by Sir George Biddell Airy \cite{Airy}, Lord Rayleigh  \cite{Rayleigh} and Sir Chandrasekhara Venkata Raman \cite{Raman}. This effect appears when waves propagate inside a bounded area and waves with certain frequencies propagate along the boundary. Such behavior for wave propagation appears in different areas and models of physics. This effect appears in acoustical three-dimensional problems with complex  geometry \cite{KatsPetr19, PetrAnt20}. In optics, the whispering gallery modes appear in the three-dimensional microresonators of different forms \cite{OptWGM1, OptWGM2, OptWGM_Arxiv, OptWGM_wedge_1, OptWGM_wedge_2}. The toroidal resonators play a very important role \cite{OptWGM2, OptWGM_Arxiv}, but due to the difficulties in production, usually in such resonators the cross-section is some convex area, which is not necessary the perfect round disc. On the other hand, the different shapes of such microresonators lead to other important properties; for example, the resonators with wedges were studied in \cite{OptWGM_wedge_1, OptWGM_wedge_2}.

The main approach in studying the whispering gallery  effect leads to the eigenproblems for the Laplace operator or Helmholz operator \cite{KellerRub60, BabBul72, Kirpich79, Laz67} inside some area with a smooth and convex boundary. It is known that the localized near the boundary eigenfunctions of the Laplace operator correspond to large eigenvalues. Stated in this way, the eigenproblem for the Laplace operator can be studied  using a semiclassical approach. It is known \cite{Laz93, Laz88, Arnold} that this approach allows one to approximate the discrete spectrum of the operator. However, it is also known that, in general case, the semiclassical function constructed in this way will not approximate the real eigenfunction, but will approximate a combination of the real eigenfunctions. For this reason, such semiclassical asymptotics are called quasimodes \cite{Arnold}. 

There are a lot of different works devoted to the study of the quasimodes for Laplacian.   We are particularly interested in the whispering gallery-type of quasimodes and, as we said, such quasimodes are localized in the neighborhood of the boundary. In the two-dimensional case, asymptotics for such functions have been fairly well studied, see, for example, \cite{KellerRub60, BabBul72, Kirpich79, Laz67, NgGreb13}.

The three-dimensional case is much more complicated comparing to the two-dimensional situation and there are not many works devoted to the study of the whispering-gallery type quasimodes for the Laplacian. A study of the localized quasimodes for the Laplacian was given mostly in the convex areas \cite{KellerRub60, Popov_G, Popov20, Popov20-1}. From the physical point of view, as we mentioned above, investigation of the localized eigenfunctions inside the solid torus \cite{OptWGM2, OptWGM_Arxiv, OptWGM_wedge_1, OptWGM_wedge_2} is also very important. In that case, the quasimodes may not be localized near the whole boundary but only in the vicinity of its part (more precisely, in some tubular neighborhood of the boundary).

Let us briefly pose the problem. Consider a domain $T$ constructed by rotating along the $z$ axis a two-dimensional domain $\Omega$ bounded by a convex curve $S$ in the plane $(x,\,z)$. We require that the boundary $\partial T$ is diffeomorphic to a torus. Inside this domain, we pose the eigenproblem for the Laplacain of finding eigenfunctions localized in the neighborhood of the boundary $\partial T$
\begin{equation}
\label{init_problem}
-\Delta u=\lambda^2 u,\,(x,\,y,\,z)\in T,\quad u|_{\partial T}=0,\quad \lambda\gg 1.
\end{equation}
We consider a problem with Dirichlet conditions; for the Neumann conditions, the reasoning is similar.

The main purpose of this paper is to construct suitable for computations asymptotic (for large values of $\lambda$) formulas of the whispering gallery-type for the problem (\ref{init_problem}).  We present such formulas with the help of semiclassical approximation and the construction of Maslov's canonical operator \cite{MF}. Moreover, the resulting formulas are written in analytical form with the help of the Airy functions, which significantly simplifies the calculations.

Since the domain $T$ is a body of revolution, the problem (\ref{init_problem}) can be reduced to a two-dimensional problem inside the domain $\Omega$. Due to the specifics of the problem statement, it contains several different scales \cite{BabBul72, Popov_G, Popov20} related to the region of localization of eigenfunctions. With such a difference in scales, one can use the idea of adiabatic approximation \cite{Peierls}. In our case, we use this method in the form of operator separation of variables, that was initially introduced in \cite{Dobr83} for the Cauchy-Poisson problem for the surface waves propagation. Later this method was implemented in the problems with different scales, for example, in the homogenization problems \cite{GrDobrSergT16, Serg22, BrGrDobr12} and for the separation of independent variables  \cite{DobrMinShl18, DobrMinNeiSh19}. 

This approach allows us to construct quasimodes for the problem (\ref{init_problem}) with any given accuracy as a formal power series of a small parameter $\lambda^{-1/3}$. 	This small parameter corresponds to the whispering gallery-type asymptotics \cite{BabBul72, Popov20}, and  we present an algorithm for constructing the terms of this power series, but we restrict ourselves in this paper to the first few terms of such a series.

In our case, we can reduce the  two-dimensional problem to two one-dimensional problems with the help of the operator separation of variables. Along the normal vector to the boundary $S$, similar to \cite{BabBul72, Kirpich79, Laz67, Popov20}, we obtain the Airy equation \cite{Olver}. The second equation, along the curve $S$, is a spectral problem for the perturbed stationary Schrödinger equation. The asymptotics of the solution to such equation are very well known (see, for example, \cite{Olver58, Erdelyi}). Here we use the results of \cite{ADNTs} because of the more explicit formulas. Note that in \cite{BabBul72, Popov20} the non-stationary type Schrödinger equation was  obtained, and this equation relates the asymptotics constructed along the geodesic and geometric parameters of this geodesic in three-dimensional convex area.

The present work is organized as follows. In Section \ref{sec_Problem_St}, we describe in detail the problem statement and the domain $T$. In the same section, we derive the two-dimensional equation and formulate a two-dimensional problem. In Section \ref{sec_Result}, we formulate the one-dimensional Schr\"{o}dinger equation and present the main theorems and results obtained in this paper. We present an explicit representation for the first few terms in power series for quasimodes with the help of Airy functions. In Subsection \ref{sec_bill}, we provide a short description of Hamilton system corresponding to the quasimodes. In Section \ref{sec_exampl}, we illustrate the results obtained in Section \ref{sec_Result}. Following ideas \cite{OptWGM_wedge_1, OptWGM_wedge_2, Arnold} we choose the region $\Omega$ as an equilateral triangle with rounded wedges. In Subsection \ref{subsec_simpl}, we write out a representation for the asymptotic of the one-dimensional Schr\"{o}dinger equation with the help of the parabolic cylinder functions \cite{Olver}.  In Section \ref{sec_sep_var}, we  describe the procedure for operator separation of variables and provide an algorithm for constructing a quasimode for any given accuracy. In Section \ref{sec_2D_eq_vyvod}, we provide a detailed derivation of the two-dimensional equation.

Authors thank S.Yu. Dobrokhotov, A.P. Kiselev, M. Sumetsky, A.Yu. Anikin and V.A. Kibkalo for valuable comments and suggestions during the preparation of the article.

This work was supported by RSF grant 22-71-10106. Authors declare no conflict of interests.

\section{Problem statement and derivation of the main equation.}
\label{sec_Problem_St}
\subsection{Three-dimensional problem statement.}
We consider the problem for the asymptotic eigenfunctions of the Laplace operator with Dirichlet boundary conditions for large values of eigenvalues inside the domain $T$ with boundary $\partial T$ diffeomorphic to the torus 
\begin{equation}
\label{Laplace_eig}
-\Delta u(x,\,y,\,z)=\lambda^2 u(x,\,y,\,z),\quad  (x,\,y,\,z)\in T,\quad u_{|\partial T}=0.
\end{equation}
Here $\Delta u=u_{xx}+u_{yy}+u_{zz}$, and $(x,\,y,\,z)$ are 3D Cartesian coordinates.

For $\lambda\gg 1$ the problem (\ref{Laplace_eig}) can be studied using a semiclassical approach which leads to the appearance of {\it quasimodes}. 

{\bf Definition 1.} We call the sequence of pairs $(u_n,\,\lambda_n^2)_{n\in\mathbb N^3}$ {\it quasimodes of order $\nu$} inside $T$ for the problem (\ref{Laplace_eig}) if
\begin{equation}
\label{quasi_def}
\|u_n\|_{L^2(T)}  = 1+o(1),\quad   u_n|_{\partial T}=0, \quad \|-\Delta u_n - \lambda_n^2 u_n \|_{L^2(T)} = O(\lambda_n^{2-\nu}), \quad {\rm as}\;\; n\to\infty.
\end{equation}
We call sequence of functions $(u_n)_{n\in\mathbb N^3}$ {\it asymptotic eigenfuncions} and the sequence of numbers $(\lambda_n^2)_{n\in\mathbb N^3}$ {\it asymptotic eigenvalues}.

It is known \cite{Laz93, Laz88, Arnold} that the numbers $\lambda_{n}^2$ approximate the eigenvalues $\lambda^2$ of the problem (\ref{Laplace_eig}). It is also known that the functions $u_{n}$ will only approximate a linear combination of true eigenfunctions.   The following proposition \cite{Laz88} holds.

{\bf Proposition 1.} If $(\lambda_n^2)_{n\in\mathbb N^3}$ is a series of asymptotic eigenvalues, then there exists a series of eigenvalues $(\tilde \lambda_n^2)_{n\in\mathbb N^3}$ of (\ref{Laplace_eig}) such that 
$$
\tilde \lambda_n^2 = \lambda_n^2 (1 + O(\lambda_n^{-\nu})).
$$

\begin{figure}[h]
\begin{center}
\includegraphics[scale=1]{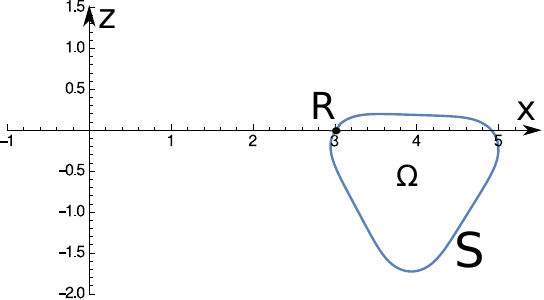}
\end{center}
\caption{Domain $\Omega$ bounded by the curve $S$ in the $(x,\,z)$ plane. Domain $\Omega$ is then rotated around the $z$ axis, along a circle of radius $R=3$. \label{pic1}}
\end{figure}

\begin{figure}[h]
\begin{center}
\includegraphics[scale=0.4]{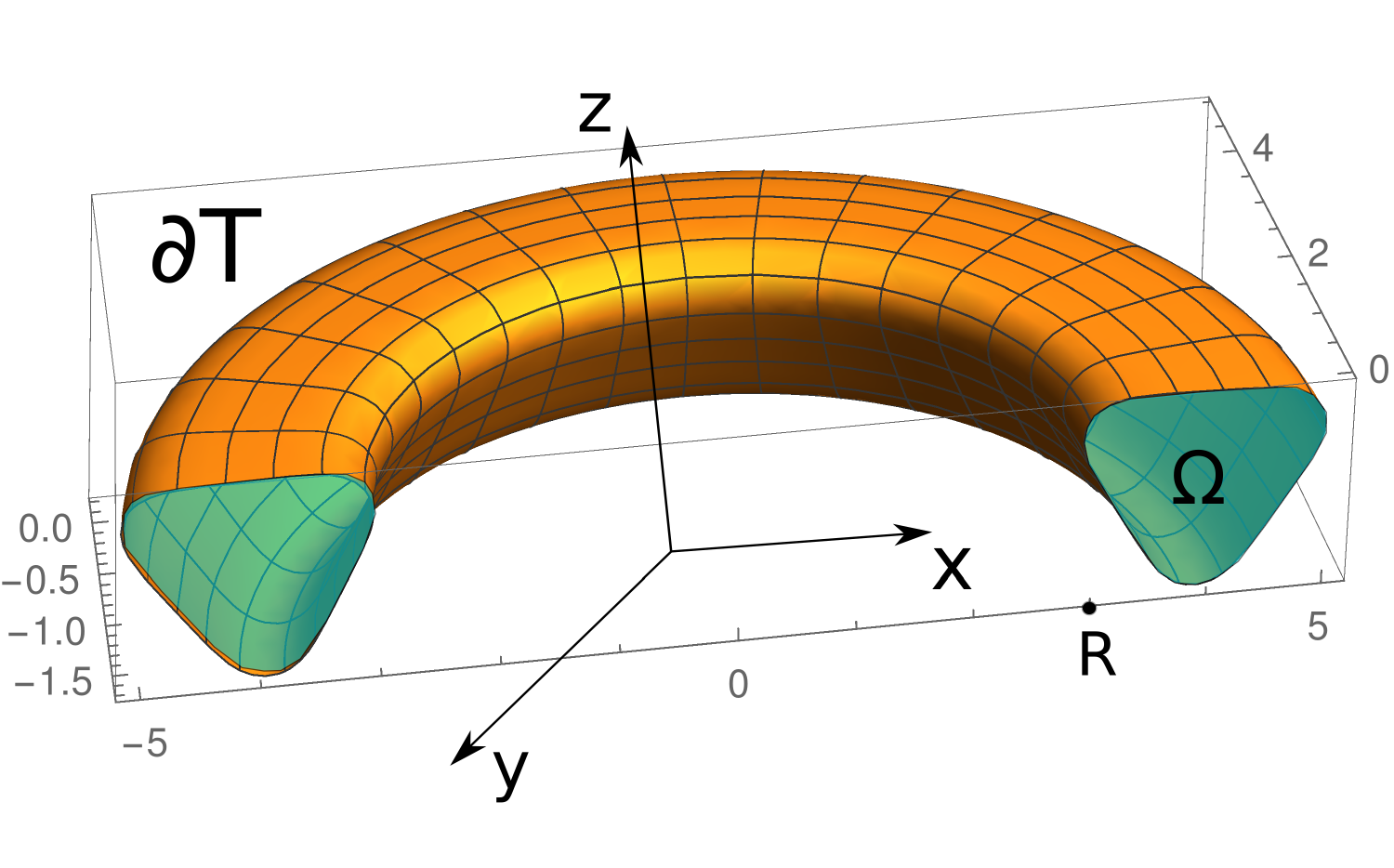}
\end{center}
\caption{Part of the surface $\partial T$ in 3D space and the domain $\Omega$. We also marked on the $x$ axis the value $R$ of the radius of the circle along which the $\Omega$ is rotated. \label{pic2}}
\end{figure}

Now let us describe the domain $T$ and formulate some conditions that we impose on it. In a three-dimensional space with Cartesian coordinates $(x,\,y,\,z)$ in the plane $(x,\,z)$ a smooth convex closed curve $S$ bounding some domain $\Omega$ is given. Then we rotate this around the $z$ axis  in the way that after a complete rotation the boundary $\partial T$ is a smooth surface and diffeomorphic to a torus.
On fig. \ref{pic1} the domain $\Omega$ and the curve $S$ in the plane $(x,\,z)$ are shown, and on fig. \ref{pic2} a part of the surface $\partial T$ is demonstrated.


\subsection{Two-dimensional problem. }
Let us pass in the equation (\ref{Laplace_eig}) to curvilinear coordinates in the vicinity of the surface $\partial T$. 
To begin with, we introduce curvilinear coordinates in the region $\Omega$ in a neighborhood of the curve $S$. Suppose that in the plane $(x,\,z)$ the curve $S$ is described by the equations
$$
x=R+Q_1(s),\quad z=Q_2(s),
$$
where $s$ is the arclength parameter on the curve and it is measured from some point $s_0$; and constant $R>0$ is large enough such that $S$ does not intersect $z$-axis (see fig.  \ref{pic1}). The functions $Q_{1,\,2}(s)$ are periodic with period $L$, where $L$ is the length of the curve $S$. We assume that the path tracing on the curve $S$ is counterclockwise.

\begin{figure}[h]
\begin{center}
\includegraphics[scale=1]{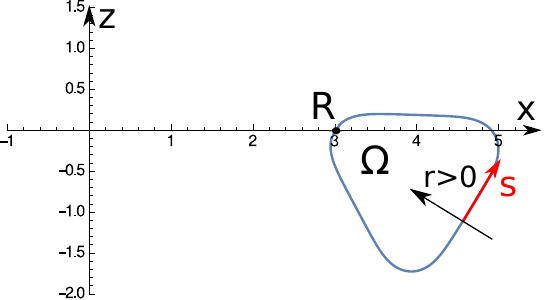}
\end{center}
\caption{Coordinates of $(r,\,s)$ near the $\Omega$ boundary. \label{pic3}}
\end{figure}

Next, we introduce the coordinate $r$ directed along the inner normal of the curve $S$ at the point $s$, so that $r>0$ inside $\Omega$, $r=0$ on the curve, and $r<0$ in the outer areas. On fig. \ref{pic3} we have schematically depicted this coordinate system.

Since the solid torus $T$ is a body of revolution, in a neighborhood of the surface $\partial T$ one can introduce the coordinate system $(r,\,s,\,\alpha)$ as follows
\begin{gather}
\label{coord_sys_bound_3D}
x(r,\,s,\,\alpha)=X(r,\,s)\sin\alpha,\quad y(r,\,s,\,\alpha)=X(r,\,s )\cos\alpha,\quad z(r,\,s,\,\alpha)=Z(r,\,s),\\
\label{coord_sys_bound}
X(r,\,s)=(R+Q_1(s))-r Q_2'(s),\quad Z(r,\,s)=Q_2(s)+r Q_1'(s),
\end{gather}
where $\alpha$ is the angle of rotation of $\Omega$ along the $z$ axis along a circle of radius $R$. Here we took into account that the inner normal to the curve $S$ is described by the vector $(-Q_2'(s),\,Q_1'(s))$.

Due to the fact that the equation (\ref{Laplace_eig}) is considered inside of a body of revolution, we can separate variables in the coordinates $(r,\,s,\,\alpha)$ and reduce the equation to two-dimensional one in the domain $\Omega$.

We introduce a new unknown function $w(r,\,s)$ as follows
\begin{equation}
\label{new_func}
u(r,\,s,\,\alpha)=\frac{e^{in \alpha}}{\sqrt{|J(r,\,s)|}}w(r,\,s), \quad n\in\mathbb{Z},
\end{equation}
where 
\begin{equation}
\label{Jac}
J(r,\,s)=(1-rk(s))X(r,\,s)
\end{equation}
is the Jacobian of the transition from Cartesian coordinates $(x ,\,y,\,z)$ to the coordinates $(r,\,s,\,\alpha)$, and by $k(s)$ we denoted the curvature of the curve $S$ at the point $s$.

Since in (\ref{Laplace_eig}), eigenvalues $\lambda^2$ are big,  it is convenient for us to  introduce a small parameter $\varepsilon\ll 1$ and operator $\hat{p}_s$  such that
$$
\lambda=\frac{\mathcal{E}}{\varepsilon},\quad \varepsilon\ll1,\quad \mathcal{E}=O(1),\quad \hat{p}_s=-i\varepsilon\frac{\partial}{\partial s}.
$$

After multiplying the equation (\ref{Laplace_eig}) by $\varepsilon^2$ and passing to the coordinates $(r,\,s,\,\alpha)$, we obtain the following two-dimensional problem with an accuracy of $O(\varepsilon^2)$ for the function $w(r,\,s)$
\begin{equation}
\label{2D_eq}
\Delta_2 w\equiv \left(-\varepsilon^2\frac{\partial^2}{\partial r^2}+\hat{p}_s\frac{1}{(1-rk(s))^2}\hat{p}_s+ \frac{a_n^2}{X^2(r,\,s)}\right)w=\mathcal{E}^2 w+O(\varepsilon^2) ,\quad w|_{r=0}=0.
\end{equation}
The correction $O(\varepsilon^2)$ is a smooth bounded function of the given order and does not contain any derivatives of the function $w$. For a more detailed derivation of the equation (\ref{2D_eq}), see section \ref{sec_2D_eq_vyvod}.

The number $a_n=n\varepsilon$ is obtained after separating the variable $\alpha$. With respect to the quantities $n$ and $\varepsilon$, we assume that the following relation holds
$$
\frac{a_n}{R}\equiv \frac{n\varepsilon}{R}=O(1).
$$
This condition means that the term $a_n^2/X^2(r,\,s)$ will remain in the main part of the equation (\ref{2D_eq}).

Equation (\ref{2D_eq}) is valid only in the area $\Pi_{m}=\{r< 1/k(s),\,s\in[0,\,L]\}$  because the coordinates $(r,\,s)$ according to (\ref{coord_sys_bound_3D}), (\ref{coord_sys_bound}) are valid only in $\Pi_m$. From the other point of view, we are interested in the asymptotics  which is localized near the  $r=0$. 
Similar to the ideas of \cite{NgGreb13, DNS_UMS}, we call a function $f(r,\,s)$ localized near $r=0$, if outside of some small area $P_{loc}\subset\Pi_{m}$ of $r=0$ the norm  $\|f\|_{L_2(\Pi_m\backslash \Pi_{loc})}$ is exponentially small. 

{\bf Remark 1.} We do not provide here the definition of the localized function, since it is very complicated and requires a lot of additional information. In our case, asymptotics are described with the help of Airy functions, which have the exponential decaying asymptotic, and which we use in our considirations.

We are looking for the localized near the curve $S$ solutions to the equation (\ref{2D_eq}). The curve $S$ is described by the equation $r=0$ and we assume that along the variable $r$ the area of the localization can be described by the small parameter $h\ll 1$ such that in this area $r=O(h)$.   
Following \cite{BabBul72, Kirpich79, Laz67}, we assume the following relation between $h$ and $\varepsilon$
\begin{equation}
\label{h_eps}
h=\varepsilon^{2/3}.
\end{equation}

We want to construct the localized asymptotic solution for the equation (\ref{2D_eq}) with precision $O(h^2)$. In that case it is convenient to introduce a new normalized variable $\rho=r/h$, $\rho=O(1)$.
We multiply the equation (\ref{2D_eq}) by $(1-h \rho k(s))^2$ and move everything to the left hand side of the equation. After that, we expand the coefficients of the equation with respect to the power of small parameter $h$ up to the $O(h^2)$. 

Let us introduce the following functions
\begin{equation}
\label{potential_A_coef}
V(s;\, \mathcal{E}^2)=\frac{a_n^2}{X^2(0,\, s)}-\mathcal{E}^2,\quad
\mathcal{A}(s;\, \mathcal{E}^2)=\left(\frac{2 a_n^2}{X^3(0,\,s)}Q_2'(s)-2 k(s)V(s;\,\mathcal{E}^2) \right)^{1/3}.
\end{equation}
Function $V(s;\, \mathcal{E}^2)$ we  call the potential. 
We also need the following differential operator
\begin{equation}
\label{oper_D_G}
\hat{D}=-\frac{\partial^2}{\partial \rho^2}+\rho \mathcal{A}^3(s;\, \mathcal{E}^2).
\end{equation}

As a result of simplification of the equation (\ref{2D_eq}) up to the $O(h^2)$,  after the indicated transformations, we obtain the following $O(h^2)$-approximation to equation (\ref{2D_eq})
\begin{equation}
\label{2D_eq_new}
\hat{H}w\equiv \left(\hat{p}_s^2+V(s;\, \mathcal{E}^2)+ h\hat{D}\right)w=O(h^2),\quad w|_ {\rho=0}=0.
\end{equation}
The description of derivation of the equation is given in Section \ref{sec_2D_eq_vyvod}.

{\bf Definition 2.}  
Let us define the class $C_{0, L}^{\infty}(\Pi)$ of the smooth functions $f(\rho,\,s)$ which are periodic with respect to $s$ with period $L$, $f(0,\,s)=0$ and 
$$
\max\limits_{s\in[0,\,L]}|f(\rho,\,s)|\to 0,\quad \max\limits_{s\in[0,\,L]}\left|\left(-i\frac{\partial }{\partial \rho}\right)^\ell\left(-i\varepsilon\frac{\partial}{\partial s}\right)^q f(\rho,\,s)\right|\to 0,\quad \forall\,\ell,\,q\in \mathbb{Z}_+,\quad  \rho	\to+\infty,
$$
where the half-strip
\begin{equation}
\label{Pi}
\Pi=\{0\le s\le L,\quad 0\le \rho<+\infty\}.
\end{equation}

Our idea is the following. We want to construct function $w(\rho,\,s)\in C_{0,L}^{\infty}(\Pi)$ such that the equation (\ref{2D_eq_new}) is satisfied with $O(h^2)$, and function $w(r/h,\,s)$ will be localized near $r=0$ in some small area $P_{loc}$. Then, similar to \cite{Laz67}, we define the smooth cutoff function $\theta(r,\,s)$, which equals to unity in $P_{loc}$ with compact support  $\text{supp}\,\theta\subset\Pi_{m}$. Thus, the function $\theta(r,\,s)w(r/h,\,s)$ will be  well defined for all $r\ge 0$ and it will still satisfy to the equation (\ref{2D_eq}) with precision $O(h^2)$.

{\bf Remark 2.} We have reduced the study of the asymptotics of the equation (\ref{2D_eq}) to a rather simple spectral problem (\ref{2D_eq_new}).
Note that in the main term, the equation (\ref{2D_eq_new}) is a one-dimensional spectral problem for the Schrödinger equation with respect to the variable $s$
$$
\left(\hat{p}_s^2+V(s;\, \mathcal{E}^2)\right)w=0.
$$
Thus, the results of \cite{MF, ADNTs} for the semiclassical asymptotics for the Schr\"{o}dinger equation can be used  to construct the asymptotics of the problem (\ref{2D_eq_new}).

{\bf Remark 3.} The correction $O(h^2)$ in (\ref{2D_eq_new}), generally speaking, is the result of action of some differential operator with smooth coefficients of order $O(h^2)$ on function $w(\rho,\,s)$. Since this functions belongs $ C_{0,L}^{\infty}(\Pi)$, the result of this action is also smooth function of  order $O(h^2)$.

Laplacian $(-\Delta)$ inside the solid torus $T$ with Dirichlet boundary condition on the surface $\partial T$ is a symmetric and positive definite operator. We impose the condition of symmetry and positive definition for the operator $\hat{H}$ with respect to the inner product $L_2(\Pi)$ acting on the functions from $C_{0,L}^{\infty}(\Pi)$. The condition 
\begin{equation}
\label{A_ineq}
\mathcal{A}(s;\, \mathcal{E}^2)>0
\end{equation}
is sufficient for positiveness of the operator $\hat{H}$, since this function is in its definition (\ref{oper_D_G}), (\ref{2D_eq_new}). 

Condition (\ref{A_ineq}) plays an important role in our considirations. It calls  the first approximation stability condition as stated in \cite{BabBul72, Popov_G, Popov20} when considering the whispering gallery-type quasimodes in the vicinity of the geodesic of the stricly convex area. The very natural description of this condition can be provided in terms of the corresponding dynamical systems and billiards, and it is given in the subsection \ref{sec_bill}.

\section{Formulation of the main results.}
\label{sec_Result}

\subsection{Short description for the operator separation of variables.}
\label{sec_short_oper_separ}
Let us now give a brief description of the procedure for the operator separation of variables.
As we noted above, there are two scales in this problem, characterized by the parameters $h$ and $\varepsilon$. Following \cite{GrDobrSergT16, Serg22, BrGrDobr12}, we will seek the solution of (\ref{2D_eq_new}), function $w(\rho,\,s)$, in the form of the action of some operator $\hat{\chi}$ on the function $\psi(s)$, which  satisfies the one-dimensional spectral problem:
\begin{equation}
\label{oper_form}
w(\rho,\,s)=\chi(\rho,\,\stackrel{2}{s},\,\stackrel{1}{\hat{p}}_s;\,\varepsilon,\,h)\psi(s),\quad L(\stackrel{2}{s},\,\stackrel{1}{\hat{p}}_s;\, \mathcal{E}^2;\,\varepsilon,\,h)\psi(s)=0.
\end{equation}
Here the numbers above the operators indicate the order of action: first, the operator of differentiation $\hat{p}_s$, then --- the operator of multiplication by the variable $s$ \cite{MF, M}.
The Dirichlet boundary condition for the function $w(\rho,\,s)$ in (\ref{oper_form}) has the form
\begin{equation}
\label{chi_boundary}
\chi(0,\,\stackrel{2}{s},\,\stackrel{1}{\hat{p}}_s;\,\varepsilon,\,h)=0.
\end{equation}
The operators $\hat{\chi}$ and $\hat{L}$ are to be defined, and $\mathcal{E}^2$ is a spectral parameter.

Let us consider the function $w(\rho,\,s)$ in the form (\ref{oper_form}) and let us impose the  condition of preserving the value of the norm (see, for example, \cite{BrGrDobr12})
$$
\|w\|^2_{L_2(\Pi)}\equiv (\hat{\chi}\psi(s),\,\hat{\chi}\psi(s))_{L_2(\Pi)}=(\psi(s),\,\hat{\chi}^*\hat{\chi}\psi(s))_{L_2(\Pi)}=(\psi(s),\,\psi(s))_{L_2[0,\,L]}\equiv \|\psi\|^2_{L_2[0,\,L]}.
$$
Here $\hat{\chi}^*$ is the adjoint operator for $\hat{\chi}$. This condition of preserving the value of the norm leads to the so-called normalization condition 
\begin{equation}
\label{normal_chi}
\int\limits_{0}^{+\infty}\hat{\chi}^*\hat{\chi}d\rho=\text{Id},
\end{equation}
where $\text{Id}$ is the identity operator.

Let us substitute (\ref{oper_form}) into equation (\ref{2D_eq_new}). The sufficient condition for the function $w(\rho,\,t)$ to be a solution for that equation, is the following operator equality 
\begin{equation}
\label{operator_equal}
\hat{H}\hat{\chi}=\hat{\chi}\hat{L}.
\end{equation}
Operators $\hat{\chi}$ and $\hat{L}$ are sought from this equation and operator $\hat{H}$ is the differential operator of the equation (\ref{2D_eq_new}). 

\begin{lemma}
\label{lm_symmetr}
Let equality (\ref{operator_equal}) hold and let operator $\hat{H}$ be symmetric and positive definite on the class of functions $C^\infty_{0,\, L}(\Pi)$ with respect to the inner product in $L_2(\Pi)$. Let the operator $\hat{\chi}$ satisfy the normalization condition (\ref{normal_chi}). 

Then operator $\hat{L}$ is at least symmetric and positive defined in the space of periodic functions from $C^\infty[0,\,L]$ with respect to the inner product in $L_2[0,\,L]$.
\end{lemma}

{\bf Proof.} Let function $w\in C^\infty_{0,\,L}(\Pi)$ be of the form (\ref{oper_form}) $w=\hat{\chi}\psi(s)$, where $\psi(s)$ is a smooth periodic function. Then we have the following equalities
$$
(\hat{H}w,\,w)_{L_2(\Pi)}=(\hat{H}\hat{\chi}\psi,\,\hat{\chi}\psi)_{L_2(\Pi)}=(\hat{\chi}\hat{L}\psi,\,\hat{\chi}\psi)_{L_2(\Pi)}=(\hat{L}\psi,\,\hat{\chi}^*\hat{\chi}\psi)_{L_2(\Pi)}=(\hat{L}\psi,\,\psi)_{L_2[0,\,L]}.
$$
We used (\ref{operator_equal}) in the second equality,  and (\ref{normal_chi}) was used in the last equality. Since operator $\hat{H}$ is positive defined, then the operator $\hat{L}$ is positive defined too. The symmetry of the operator $\hat{L}$ is also following from these equalities.
{\bf Lemma \ref{lm_symmetr} is proven.}

Suppose, that for symbols of operators $\hat{\chi}$ and $\hat{L}$ the following expansions  are valid
\begin{gather*}
\chi(\rho,\,s,\,p_s;\mathcal{E}^2;\varepsilon,\,h)=\chi_0(\rho,\,s,\,p_s;\,\mathcal{E}^2; h)-i\varepsilon \chi_1(\rho, \,s,\,p_s;\,\mathcal{E}^2; h)+O(\varepsilon^2),\\
L(s,\,p_s;\mathcal{E}^2;\,\varepsilon,\,h)=L_0(s,\,p_s;\, \mathcal{E}^2;\, h)-i\varepsilon L_1(s,\,p_s;\, \mathcal{E}^2;\, h)+O(\varepsilon ^2).
\end{gather*}
In accordance to the scaling (\ref{h_eps}) between small parameters $\varepsilon$ and $h$, we assume that the following expansions with respect to the small parameter $\sqrt{h}=\varepsilon^{1/3}$ are hold
\begin{gather*}
L_0(s,\,p_s;\, \mathcal{E}^2;\, h)=L_0^0(s,\,p_s;\, \mathcal{E}^2)+h L_0^2(s,\,p_s;\, \mathcal{E}^2),\\
\chi_0(\rho,\,s,\,p_s;\,\mathcal{E}^2; h)=\chi_0^0(\rho,\,s,\,p_s; \mathcal{E}^2)+\sqrt{h}\chi_0^1(\rho,\,s,\,p_s; \mathcal{E}^2)+O(h),
\end{gather*}
and functions $L_1$ and $\chi_1$ can be presented as series with small parameter $\sqrt{h}$.
With these expansions, we can pass in (\ref{operator_equal}) to the symbols for the operators $\hat{H}\hat{\chi}$ and $\hat{\chi}\hat{L}$ as operators with respect to variable $s$. After combination of the coefficients with the same powers of $\varepsilon$, we obtain the system of equations for the coefficients of these expansions. The first two equations, which correspond to the powers $\varepsilon^0=1$ and $\varepsilon$, 	have the following form
\begin{gather}
\label{symb_equal_L0}
\Bigl(H_0(s,\,p_s)+h\hat{D}\Bigr)\chi_0(\rho,\,s,\,p_s; \mathcal{E}^2)=\chi_0(\rho,\,s,\,p_s; \mathcal{E}^2)L_0(s,\,p_s; \mathcal{E}^2),\\
\label{symb_equal_L1}
H_{0p_s}\chi_{0s}+(H_0+h\hat{D})\chi_1=\chi_0L_1+\chi_{0p_s}L_{0s}+\chi_1L_0.
\end{gather}
Here we denoted
$$
H_0(s,\,p_s)=p_s^2+V(s;\,\mathcal{E}^2)\equiv p_s^2+\frac{a_n^2}{X^2(0,\,s)}-\mathcal{E}^2,\quad 
\hat{D}=-\frac{\partial^2}{\partial \rho^2}+\rho \mathcal{A}^3(s;\, \mathcal{E}^2).
$$

Generally speaking, the procedure of calculation of the symbols of the operators can be done with an arbitrary given accuracy $O(\varepsilon^{k/3})$, $k\in\mathbb{Z}_+$.  In our case, we strict ourselves to the precision $O(h^2)=O(\varepsilon^{4/3})$. 

In order to solve the equation (\ref{2D_eq_new}) with the precision $O(h^2)$ in the $L_2(\Pi)$ norm, we need the equality (\ref{operator_equal}) to be held with precision $O(h^2)$.  In order to do so, we need to determine symbol $L$  with precision $O(h^2)$ and symbol $\chi$ with precision $O(h)$. 

Let us denote for convenience the sum $\chi_0^0+\sqrt{h}\chi_0^1$ as
$$
\mathcal{X}_0(\rho,\,s,\,p_s;\,\mathcal{E}^2)=\chi_0^0(\rho,\,s,\,p_s; \mathcal{E}^2)+\sqrt{h}\chi_0^1(\rho,\,s,\,p_s; \mathcal{E}^2),
$$
so that $\chi_0=\mathcal{X}_0+O(h)$.

\begin{statement} 
\label{st_one_dim_eq}
The following equalities are hold
\begin{gather}
\label{L_oper}
L_0(s,\,p_s;\, \mathcal{E}^2;\, h)=p_s^2+V(s;\, \mathcal{E}^2)+ht_k \mathcal{A}^2(s;\, \mathcal{E}^2),\\
\label{chi0}
\mathcal{X}_0(\rho,\,s,\,p_s)=\frac{\sqrt{\mathcal{A}(s;\, \mathcal{E}^2)}}{|Ai'(-t_k)|} Ai(-t_k+\rho \mathcal{A}(s;\, \mathcal{E}^2))\left(1-\sqrt{h}\frac{i}{2} \rho^2 \frac{\mathcal{A}'(s;\, \mathcal{E} ^2)}{\mathcal{A}(s;\, \mathcal{E}^2)} p_s\right).
\end{gather}
Here the number $(-t_k)$ is the $k$-th root of the Airy function: $Ai(-t_k)=0$ due to the boundary condition (\ref{chi_boundary}). Potential $V(s;\, \mathcal{E}^2)$ and the function $\mathcal{A}(s;\, \mathcal{E}^2)$ are defined in (\ref{potential_A_coef}). The normalization condition (\ref{normal_chi}) for (\ref{chi0}) is satisfied with the precision $O(h)$.

And, moreover, the following operator 
$$
\hat{R}\equiv \hat{H}\hat{\mathcal{X}}_0-\hat{\mathcal{X}}_0\hat{L}_0
$$
is the differential operator with respect to $s$ and $\rho$ with smooth and bounded coefficients. The symbol of the operator $\hat{R}$ satisfies the inequality
\begin{equation}
\label{oper_eq_real}
|R(s,\,p_s,\,\rho,\,\partial_\rho)|\le C_0 h (1+\sqrt{h}|p_s|)(|\partial_\rho|^2+|\partial_\rho|)+C_1 h^2(|p_s|^2+1)(1+O(h))+C_2h^2\varepsilon |p_s|,
\end{equation}
where $C_0,\,C_1,\,C_2>0$ are some constants and $p_s$ corresponds to the derivative with respect to $s$ and $\partial_\rho$ corresponds to the derivative with respect to $\rho$.

\end{statement}

{\bf Proof } of the statement \ref{st_one_dim_eq} is given in section \ref{sec_sep_var}.

\subsection{Main theorems.}
\label{sec_main_thm}

In this section, we present the main theorems that describe the form of the localized asymptotic solution to the Laplace equation (\ref{Laplace_eig}).
These theorems are based on the operator separation of variables (\ref{oper_form}) and reduction  the two-dimensional problem to the one-dimensional problem. 

\paragraph{Asymptotic solution to the one-dimensional Schr\"{o}dinger equation.}
We begin with the formulation of the results for the one-dimensional equation. Then we will formulate the theorems about the localized asymptotic solution to the equations (\ref{2D_eq_new}) and (\ref{2D_eq})  and finally  describe the asymptotic for the three-dimensional Laplace equation (\ref{Laplace_eig}). 

Following statement \ref{st_one_dim_eq} and (\ref{L_oper}), the approximate one-dimensional equation has the form
\begin{equation}
\label{1D_eq_base}
\hat{L}_0\psi(s)\equiv \biggl(\hat{p}^2_s+V(s;\,\mathcal{E}^2)+h t_k \mathcal{A}^2(s;\,\mathcal{E}^2)\biggr)\psi(s)=0.
\end{equation}
This equation has the form of the perturbed one-dimensional Schr\"{o}dinger equation, and the semiclassical asymptotic solution for it can be written based on the results \cite{MF, ADNTs}. 

We assume that the spectral parameter $\mathcal{E}^2$ in the problem (\ref{1D_eq_base}) has the following expansion in the small parameter
\begin{equation}
\label{spectr_expan}
\mathcal{E}^2=E^2+ht_k E_1+O(h^2).
\end{equation}

We will consider two situations for the equation (\ref{1D_eq_base}). The first situation is when the parameter $\mathcal{E}^2$ is big enough such the equation 
\begin{equation}
\label{turn_eq}
V(s;\, E^2)=0 \Leftrightarrow \frac{a_n^2}{X^2(0,\,s)}-E^2=0
\end{equation} 
does not have any solution in $s\in[0,\,L]$. This situation is called the absence of the turning points and the asymptotics for the equation has form of the WKB exponential. In that case, the first approximation condition (\ref{A_ineq}) has to be true for all $s\in[0,\,L]$. 

The second situation is when the parameter $E^2$ is such that the equation (\ref{turn_eq}) has two solutions $s_{\pm}$ (we assume $s_{-}<s_{+}$). These points are called the turning points. In that case the condition (\ref{A_ineq}) has to be true for $s\in[s_{-},\, s_{+}]$.

It is also known that the solution to the spectral Schr\"{o}dinger equation does not exist for any value of spectral parameter $\mathcal{E}^2$. The quantisation rules select the numerable set of the values when the solution exists. In the case of absence of the turning points, the quantisation rule is just the condition of periodicity of the solution. The case with turning points is more complicated and it leads to the Bohr-Sommerfeld quantization rule \cite{MF, ADNTs}.

Let us first describe the {\bf WKB asymptotic} for the equation (\ref{1D_eq_base}).

\begin{statement}
\label{st1_as_1D_WKB}
Let us fix positive integer numbers $n$, $k$ and $m$ and let the spectral parameter of the problem (\ref{1D_eq_base}) has the form
$$
\mathcal{E}^2_{n,k,m}=E^2_{n,\,m}+h t_k E_1+O(h^2).
$$

We assume that there are no turning points and the condition $V(s;\, E^2_{n,\,m})<0$ and the first approximation condition (\ref{A_ineq}) are held for all $s\in[0,\,L]$. Let $E^2_{n,\,m}$ satisfies the quantization rule
\begin{equation}
\label{E_period}
\int\limits_{0}^{L}\sqrt{|V(\sigma;\, E_{n,\,m}^2)|}d\sigma=2\pi m\varepsilon,\quad m\in\mathbb{Z}_ +.
\end{equation}
The correction $E_1$ is defined as follows
\begin{equation}
\label{E1_period}
E_1=\langle \mathcal{A}^2\rangle_{L}\equiv \left(\int\limits_{0}^{L}\frac{ds}{\sqrt{|V( s;\, E_{n,\,m}^2)|}}\right)^{-1}\left(\int\limits_{0}^{L}\frac{\mathcal{A}^2(s;\, E_{n,\,m}^2)}{\sqrt{V(s;\, E_{n,\,m}^2)}}ds\right).
\end{equation}

Let us define the following WKB exponential 
\begin{gather}
\label{WKB_as}
\psi(s)=A_0\sqrt[4]{\frac{V(0;\, E_{n,m}^2)}{V(s;\, E_{n,m}^2)}}
exp\left(\frac{i}{\varepsilon}\left[\int\limits_{0}^{s}\sqrt{|V(\sigma;\, E_{n,m}^2)|}d\sigma+\frac{h t_k}{2}\int\limits_{0}^{s}\frac{\langle \mathcal{A}^2\rangle_{L}-\mathcal{A}^2(\sigma;\, E_{n,m}^2)}{\sqrt{|V(\sigma;\, E_{n,m}^2)|}}d\sigma\right]\right),
\end{gather} 
where constant $A_0$ is chosen in a way, that $\|\psi(s)\|_{L_2[0,\,L]}=1$. 

Then the following estimate is valid
$$
\|\hat{L}_0\psi(s)\|_{L_2[0,\,L]}=O(h^2).
$$
\end{statement}

{\bf Proof.}
This statement can be simply verified by the substitution into the equation (\ref{1D_eq_base}) of the WKB ansatz $\psi(s)=A(s)exp(i \Phi(s)/\varepsilon)$. After that  all is left, is to present functions $A(s)$ and $\Phi(s)$ as a series of $h$. {\bf Statement \ref{st1_as_1D_WKB} is proven. }

{\bf Case with the turning points. } Let us now turn our attention to the situation when the Schrödinger equation has turning points. As we mentioned above, we consider the situation of two turning points $s_\pm$, which are the solution to the equation (\ref{turn_eq}).
For definiteness, we assume that $s_-<s_+$.

The values of the spectral parameter are determined from the quantization rule, which leads to the transcendental equation for $E^2$
\begin{equation}
\label{Bohr_Zommerfeld}
\frac{1}{\varepsilon\pi}\int\limits_{s_-}^{s_+}\sqrt{|V(\sigma;\, E^2)|}ds=m+\frac{1}{2},\quad m\in\mathbb{Z}_+.
\end{equation}
This equation is the Bohr-Sommerfeld quantization rule.

The correction $E_1$ in (\ref{spectr_expan}) is equal to
\begin{equation}
\label{add_E}
E_1=\langle\mathcal{A}^2 \rangle_{T} \equiv
\left(\int\limits_{s_-}^{s_+}\frac{ds}{\sqrt{|V(s;\, E^2)|}}\right)^{-1}\left(\int \limits_{s_-}^{s_+}\frac{\mathcal{A}^2(s;\, E^2)ds}{\sqrt{|V(s;\, E^2)|}}\right).
\end{equation}

Let us pass to the description of the asymptotic function for (\ref{1D_eq_base}) in that case. It is known that the asymptotics of the Schrödinger equation is described by the Airy function $Ai(x)$ in the vicinity of the turning point \cite{Olver58, Erdelyi, ADNTs}. Due to the presence of corrections for the potential in the equation (\ref{1D_eq_base}), in our case, the asymptotics will also include the derivative of the Airy function. 

Let us define two functions
\begin{gather}
\label{psi_pm}
\psi_\pm(s;\, E^2)=\sqrt{2\pi}e^{\pm i\frac{\pi}{4}}\kappa^\pm_m\frac{1}{\sqrt [4]{|V(s;\, E^2)|}}\times\\
\nonumber
\times\Biggl[
\left(\frac{3}{2\varepsilon}\right)^{1/6}a^{ev}_\pm(s)Ai\left(-\left(\frac{3}{2\varepsilon }\right)^{2/3}\Phi_\pm(s)\right)\mp\left(\frac{3}{2\varepsilon}\right)^{-1/6}a^{odd} _\pm(s)Ai'\left(-\left(\frac{3}{2\varepsilon}\right)^{2/3}\Phi_\pm(s)\right)\Biggr].
\end{gather}
Here the numbers are $\kappa^+_m=1$ and $\kappa^-_m=(-1)^m$, and the argument (phase) of the Airy function and its derivative is defined as follows
$$
\Phi_{\pm}(s;\, E^2)=\left(\int\limits_{s_\pm}^{s}\sqrt{|V(\sigma;\, E^2)|}d \sigma\right)^{2/3}.
$$
The corresponding amplitudes have the form
$$
a^{ev}_\pm(s)=\sqrt[4]{|\Phi_\pm(s;\, E^2)|}\cos(g_1(s;\, E^2)),\quad a^{odd}_{\pm}=\frac{\sin(g_1(s;\, E^2))}{\sqrt[4]{|\Phi_{\pm}(s;\, E ^2)|}},
$$
where
$$
g_1(s;\, E^2)=\frac{h}{\varepsilon}\frac{t_k}{2}\int\limits_{s}^{s_+}\frac{\langle\mathcal{A}^2 \rangle_{T} -\mathcal{A}^2 (s;\, E^2) }{\sqrt{|V(s;\, E^2)|}}ds .
$$

The function $g_1(s;\, E^2)$ enters the asymptotics as an argument of trigonometric functions. A comparison with the WKB exponential (\ref{WKB_as}) shows that this function thus determines the phase correction at $h/\varepsilon$.

The functions $\psi_{\pm}(s)$ are defined inside the interval $s\in(s_{-},\,s_ {+})$. Let us choose some small $\delta>0$ and extend them in $\delta$-vicinity beyond this interval similar to \cite{ADNTs}. Namely, we define the extension of some function $f(s)$ beyond the points $s_{\pm}$ as follows
\begin{equation}
\label{continue}
f^{\pm}(s)=\sum\limits_{j=0}^{\ell-1}C_\ell^{j-1}(-1)^jf(s_{\pm}+j(s_{\pm}-s)).
\end{equation}
Here $\ell\in\mathbb{N}$ is some number (in practical calculations we used $\ell=3$), and  $f^{+}(s)$ is an extension to the set $s\ge s_{+}$, and the function $f^{-}(s)$ is an extension to $s\le s_{-}$. 

Since we extend Airy function and its derivative which exponentially decaying outside interval $(s_{-}-\delta,\,s_{+}+\delta)$, we multiply functions $\psi_{\pm}(s)$ on the smooth cutoff function $\zeta(s)$ with compact support and which equals unity inside this interval. This procedure gives us the smooth fucntion on the whole interval $s\in[0,\,L]$.

\begin{statement}
\label{st1_as_1D}
Let us fix positive integer numbers $n$, $k$ and $m$ and let the spectral parameter of the problem (\ref{1D_eq_base}) has the form
$$
\mathcal{E}^2_{n,k,m}=E^2_{n,m}+h t_k E_1+O(h^2).
$$
Suppose that the equation (\ref{1D_eq_base}) has turning points $s_{\pm}$ and the value $E^2_{n,m}$ is determined by (\ref{Bohr_Zommerfeld}) and the correction $E_1$ is determined by (\ref{add_E}). 

Let us fix some small number $\delta>0$ and define the function $\psi(s)$ as follows
\begin{equation}
\label{psi_as}
\psi(s)=A_0\zeta(s)\times 
\begin{cases}
\psi_{-}(s),\quad s\in[0,\,s_{+}-\delta],\\
\psi_{+}(s),\quad s\in[s_{-}+\delta,\,L],
\end{cases}
\end{equation}
where functions $\psi_\pm(s)$ are defined in (\ref{psi_pm}) and their continuation outside of the interval $(s_{-},\,s_{+})$ is given by (\ref{continue}).  The $\zeta(s)$ is smooth cutoff function with compact support and which equals unity inside the interval $(s_{-}-\delta,\,s_{+}+\delta)$. The constant $A_0$ is chosen in such way that 
$$
\|\psi(s)\|_{L_2[0,\,L]}=1.
$$

Then the  function $\psi(s)$ satisfies the condition
$$
\|\hat{L}_0\psi(s)\|_{L_2[0,\,L]}=O(h^2).
$$

\end{statement}

{\bf Proof} of the statement \ref{st1_as_1D} almost verbatim repeats the results of \cite{MF, ADNTs} and the example given in \cite{ADNTs} for the one-dimensional Schrödinger equation with a small correction. For this reason, we do not provide a substantiation for this assertion here. We only note the following point, which distinguishes our situation from that considered in \cite{ADNTs}. In our case, there is an additional small parameter $h$, which significantly affects the accuracy of the constructed asymptotics, and the corresponding formulas are given with an accuracy of $O(h^2)$, in contrast to \cite{ADNTs}, where there is only one parameter $\varepsilon$, and the asymptotics is constructed with an accuracy of $O(\varepsilon^2 )$. Note in addition, that the formula (\ref{psi_as}) is the result of the simplifications of the Maslov's Canonical Operator \cite{MF}.
{\bf Statement \ref{st1_as_1D} is proven. }

{\bf Remark 4.} Functions $\psi_{\pm}(s)$ are coincide up to the $O(h^2)$ on the interval $(s_-+\delta,\, s_+-\delta)$ outside the turning points. One should chose one of these functions in (\ref{psi_as}) on this interval.

{\bf Remark 5.} Formula (\ref{psi_as}) provides the so-called ``uniform'' asymptotics for the equation (\ref{1D_eq_base}), in the sense that representations in the terms of Airy functions are valid on the sufficiently large vicinities of the turning points.




{\bf Remark 6.} If we suppose that $|s_+-s_-|$ is sufficiently small, then instead of (\ref{psi_as}), we can describe  the asymptotic solution to the equation (\ref{1D_eq_base})  with parabolic cylinder functions \cite{Olver} uniformly for all values of the $s$. This simplification is given in the subsection \ref{subsec_simpl}.

\paragraph{Asymptotic solutions of the two-dimensional equations. }
After we obtained the function  $\psi(s)$,  we can construct the asymptotics for the two-dimensional equation (\ref{2D_eq_new}) using the operator separation of variables (\ref{oper_form}).  In order to do so, we need to act with the operator $\hat{\chi}$ on the function $\psi(s)$.

\begin{theorem}
\label{thm1_base}
Let us fix positive integers $(n,\,k,\,m)$ and define  the pair $(\psi(s),\,\mathcal{E}_{n,k,m}^2)$  according to the statements \ref{st1_as_1D_WKB} or \ref{st1_as_1D}.

Then the function
\begin{equation}
\label{w_main_sol}
w(\rho,\,s)=\frac{\sqrt{\mathcal{A}(s;\, \mathcal{E}_{n,k,m}^2)}}{|Ai'(-t_k)|}  Ai(-t_k+\rho \mathcal{A}(s;\, \mathcal{E}_{n,k,m}^2))\left(1-\sqrt{h}\frac{i} {2} \rho^2 \frac{\mathcal{A}'(s;\, \mathcal{E}_{n,k,m}^2)}{\mathcal{A}(s;\, \mathcal{E}_{n,k,m}^2)} \hat{p}_s\right)\psi(s)
\end{equation}
belongs the space $C_{0,L}^{\infty}(\Pi)$ and satisfies the  equality 
\begin{equation}
\label{as_2D_norm}
\|\hat{H}w\|_{L_2(\Pi)}=O(h^2),
\end{equation} 
where $\Pi$ is the half-strip (\ref{Pi}) and the operator $\hat{H}$ is the operator of the equation (\ref{2D_eq_new}).

\end{theorem}

{\bf Proof of the theorem \ref{thm1_base}. } First of all, notice that function (\ref{w_main_sol}) has the form $w(\rho,\,s)=\hat{\chi}_0\psi(s)$, where $\hat{\chi}_0$ is defined in (\ref{chi0}). Then with the help of statement \ref{st_one_dim_eq} and (\ref{oper_eq_real}), we can write down the equalities
$$
\|\hat{H}w(\rho,\,s)\|_{L_2(\Pi)}=\|\hat{\chi}_0\hat{L}_0 \psi(s)+\hat{R}\psi(s)\|_{L_2(\Pi)}\le\|\hat{L}_0\psi\|_{L_2[0,\,L]}+\|\hat{R}\psi(s)\|_{L_2[0,\,L]}=O(h^2).
$$ 
The estimate $\|\hat{R}\psi(s)\|_{L_2[0,\,L]}=O(h^2)$ is valid because the estimate (\ref{oper_eq_real}) for the symbol of $\hat{R}$, the fact that, $\psi(s)$ does not depend on $\rho$, and boundness  of the derivative $\hat{p}_s^2\psi(s)$. The estimate $\|\hat{L}_0\psi\|_{L_2[0,\,L]}=O(h^2)$ is valid because the statements \ref{st1_as_1D_WKB} and \ref{st1_as_1D}.

The fact of the inclusion $w(\rho,\,s)\in C_{0,L}^{\infty}(\Pi)$ is the concequence of the fact that function Airy $Ai(-t_k+\rho\mathcal{A}(s;\, \mathcal{E}^2))$ exponentially decaying with derivatives when $\rho\to+\infty$.
{\bf Theorem \ref{thm1_base} is proven.}

{\bf Remark 7.} Due to the statement \ref{st_one_dim_eq} and normalization condition (\ref{normal_chi})  for the operator $\hat{\chi}_0$, we have the following estimation
$$
\|w\|_{L_2(\Pi)}=1+O(h).
$$

Let us recall now, that generally speaking, we are interested in the localized near $r=0$ solution of the two-dimensional equation (\ref{2D_eq}). 

Let $C>0$, $C=O(1)$ be some constant and let us define area $\Pi_{loc}$ such that 
$$
\Pi_{loc}\equiv\left\{s\in[0,\,L],\quad  r\in\left[0,\,\sqrt{h}\frac{C}{k(s)}\right]\right\}\subset\Pi_m\equiv\left\{s\in[0,\,L],\quad  r\in\left[0,\,\frac{1}{k(s)}\right]\right\}.
$$
Area $\Pi_m$ is the area, where coordinates $(r,\,s)$ are valid.

{\bf Proposition 2.} Let function $w(\rho,\,s)$ be as in (\ref{w_main_sol}), then the function $w(r/h,\,s)$ is exponentially small outside the $\Pi_{loc}$.

Indeed, outside of $\Pi_{loc}$, we have $r\ge \sqrt{h} C/k(s)$, then for the argument of the Airy function in (\ref{w_main_sol}) we have the following estimate 
$$
-t_k + \mathcal{A}(s;\,\mathcal{E}^2) r/ h\ge -t_k+\frac{C}{\sqrt{h}}\frac{\mathcal{A}(s;\, \mathcal{E}^2)}{k(s)}.
$$ 
It means that argument of the Airy function is of order $O(1/\sqrt{h})$ when $h\ll1$, and thus Airy function is exponentially small.

Let us define the cutoff function $\theta(r,\,s)$ as follows. This is the smooth function in the  half-strip $[0\le s\le L]\times [0\le r<+\infty)$; it has a compact support $\text{supp}\,\theta\subset\Pi_m$ and this function equals unity inside the $\Pi_{loc}$.

Since the function $w(r/h,\,s)$ is exponentially small outside the $\Pi_{loc}$, then the function 
\begin{gather}
\label{w_func_fin}
\tilde{w}(\frac{r}{h},\,s)=\theta(r,\,s) w(\frac{r}{h},\,s)
\end{gather}
differs from $w(r/h,\,s)$ on the exponentially small correction in $\Pi_{m}$. 

\begin{theorem}
\label{cor1}
Function (\ref{w_func_fin}) satisfies the following estimation
\begin{equation}
\label{Delta_2_res}
\|(\Delta_2-\mathcal{E}_{n,k,m}^2)\tilde{w}\|_{L_2([0\le s\le L]\times [0\le r<+\infty))}=O(h^2),
\end{equation}
where $\Delta_2$ is the differential operator in the equation (\ref{2D_eq}) and $\mathcal{E}_{n,k, m}$ is as in theorem \ref{thm1_base}.

\end{theorem}

{\bf Proof.} After substitution of the function $\tilde{w}$ into the equation (\ref{2D_eq}), we have to estimate the following norm
$\|(\Delta_2-\mathcal{E}_{n,k,m}^2+O(\varepsilon^2))\tilde{w}\|$.
The correction $O(\varepsilon^2)$ appears after passing from the three-dimensional equation to two-dimensional and it does not contain any operators of differentiation. Thus, we can omit this correction because we need correction $O(h^2)$.

Therefore we can write the following sequence
\begin{gather}
\nonumber
\|(\Delta_2-\mathcal{E}_{n,k,m}^2)\tilde{w}\|_{L_2([0\le s\le L]\times [0\le r<+\infty))}\le\|\theta\,(\Delta_2-\mathcal{E}_{n,k,m}^2)w\|+C\left\|h^2 \theta_{rr}w+h2\theta_r w_r+O(\varepsilon)\right\|=\\
\label{norm_cor}
=\left\|\frac{\theta}{(1-h\rho k(s))^2}(\hat{H}+O(h^2)) w\right\|_{L_2(\Pi)}+O(h^\infty)=O(h^2).
\end{gather}
Here $C>0$ and in the first norm we passed from coordinates $(r,\,s)$ to $(\rho,\,s)$ and used the equality 
$$
(1-h\rho k(s))^2(\Delta_2-\mathcal{E}_{n,k,m}^2)=\hat{H}+O(h^2).
$$ 
This equality follows from the expansion of the coefficients in the  equation (\ref{2D_eq}) and the definition of the  operator $\Delta_2$. Full derivation of this equality is given in the section \ref{sec_2D_eq_vyvod}.  Thus from the theorem \ref{thm1_base}, we see that  first norm in (\ref{norm_cor})  is $O(h^2)$ since function $\theta(\rho,\,s)/(1-h\rho k(s))$ is correctly defined (smooth and bounded) in $\Pi$ and $\hat{H}w=O(h^2)$.

Function $w(r/h,\,s)$ is localized in $\Pi_{loc}$,  but in this area derivatives $\theta_{r}\equiv \theta_{rr}\equiv 0$, therefore the norm $\left\|h^2 \theta_{rr}w+h 2\theta_r w_r+O(\varepsilon)\right\|=O(h^\infty)$, in other words, it is exponentially small.
 The correction $O(\varepsilon)$ appears because the derivative $\hat{p}_s\theta=-i\varepsilon\theta_s=O(\varepsilon)$. 
{\bf Theorem \ref{cor1} is proven.}

{\bf Remark 8.} Due to the presence of the potential in the two-dimensional equation (\ref{2D_eq}), we can say, that the theorem \ref{cor1} is some generalization of the results from \cite{BabBul72}, because in \cite{BabBul72} the two-dimensional case was considered where only the WKB form considered along the boundary.

\paragraph{Asymptotic solution to the initial three-dimensional problem.}
Now we can formulate theorem for the localized asymptotic solution to the eigenproblem (\ref{Laplace_eig}) of the Laplacian inside the solid torus $T$. 

Let us denote $T_m$ as the area inside $T$ where the change of variables (\ref{coord_sys_bound_3D}) is valid. Under the change of variables (\ref{coord_sys_bound}),  the area $\Pi_{loc}$ corresponds to some neighborhood $\Omega_{loc}$ of $S$ inside $\Omega$. The rotation of $\Omega_{loc}$ along the $z$-axis will give the domain  where the whole three-dimensional function will be localized. It is obvious, that this domain is inside the area $T_m$.

\begin{theorem}
\label{thm_Laplace_eig}
Consider the spectral problem (\ref{Laplace_eig})
$$
-\Delta u(x,\,y,\,z)=\lambda^2 u(x,\,y,\,z),\, (x,\,y,\,z)\in T,\quad u_{|\partial T}=0,
$$
inside the solid torus $T$ and let $(r,\,s,\,\alpha)$ be the coordinates near boundary $\partial T$ according to (\ref{coord_sys_bound_3D}). 

Define the spectral parameter as follows
\begin{equation}
\label{lambda_fin}
\lambda_{n,\,k,\,m}^2=\frac{\mathcal{E}_{n,\,k,\,m}^2}{\varepsilon^2},\quad \mathcal{E}_{n,\,k,\,m}^2=E_{n,\,m}^2+\varepsilon^{2/3} t_k E_1+O(\varepsilon^{4/3}),\quad \varepsilon\ll 1,
\end{equation}
where $(n, k, m)$ are positive integers and $\mathcal{E}_{n,k, m}$ is defined as in theorem \ref{thm1_base}. Let us define function 
\begin{gather}
\label{base_asympt}
u(x,\,y,\,z)=
\begin{cases}
e^{in \alpha}\frac{\tilde{w}(r/\varepsilon^{2/3},\,s)}{\sqrt{(1-rk(s))X(r,\,s)}},\, (x,\,y,\,z)\in T_{m},\\
0,\,(x,\,y,\,z)\in T\backslash T_{m},
\end{cases}\\
\nonumber
\alpha=\alpha(x,\,y),\quad r=r(x,\,y,\,z),\quad s=s(x,\,y,\,z).
\end{gather}
Here function $\tilde{w}(r/\varepsilon^{2/3},\,s)$ is as in theorem \ref{cor1}.

Then the pair $(\lambda^2_{n,\,m,\,k},\, u(x,\,y,\,z))$ is the quasimode of order $\nu=4/3$ for the initial spectral equation (\ref{Laplace_eig}). 


\end{theorem}

{\bf Proof} of the theorem \ref{thm_Laplace_eig} follows from the theorems \ref{thm1_base} and \ref{cor1} and from the constructions described above. We give here just brief demonstration for this proof.

The following sequence of the equalities is true
$$
\|(-\Delta-\lambda^2_{n,k,m})u\|_{L_2(T)}=\frac{1}{\varepsilon^2}\|(-\varepsilon^2\Delta-\mathcal{E}^2_{n,k,m})u\|_{L_2(T)}=\frac{1}{\varepsilon^2}\|(\Delta_2-\mathcal{E}^2_{n,k,m})\tilde{w}\|_{L_2([0\le s\le L]\times [0\le r\le +\infty))}.
$$
Here in the last equality we made the change of variables (\ref{coord_sys_bound_3D}), (\ref{coord_sys_bound}) and passed to the two-dimensional equation (\ref{2D_eq}) in coordinates $(r,\,s)$.

Using theorem \ref{cor1}, we have 
$$
\|(-\Delta-\lambda^2_{n,k,m})u\|_{L_2(T)}=\frac{1}{\varepsilon^2}O(h^2)=O(\varepsilon^{-2/3})=O(\lambda_{n, k,m }^{2/3}).
$$
We obtain the order $\nu=4/3=2-2/3$ for the quasimodes from the last equality. {\bf Theorem \ref{thm_Laplace_eig} is proven.}

{\bf Remark 9.} Note that for $n\asymp m$\footnote{Recall, that for positive $f(\varepsilon)>0$, $g(\varepsilon)>0$, $f(\varepsilon)\asymp g(\varepsilon)$ as $\varepsilon\to0$ means, that for some constants $C>c>0$ the following inequality holds $c < f(\varepsilon) / g(\varepsilon) < C$. } and $k = O(1)$ as $n\to\infty$ the order of asymptotic eigenvalues is $\lambda_{n,m,k}^2 \asymp n^2$ and the distances 
$$
\delta^{(1)}_{n,m,k} = \lambda_{(n+1),m,k}^2 - \lambda_{n,m,}^2 \asymp n,\quad \delta^{(3)}_{n,m,k} = \lambda_{n,(m+1),k}^2 - \lambda_{n,m, k}^2 \asymp n
$$ 
are larger than correction $\delta = C \lambda_{n,m, k}^{2/3} \asymp n^{2/3}$.

{\bf Remark 10.} If one-dimensional equation (\ref{1D_eq_base}) has truning points, then function (\ref{base_asympt}) will be localized near some part of the surface $\partial T$. Otherwise, this function will be localized near the whole surface $\partial T$.

{\bf Remark 11.} The functions (\ref{base_asympt}) corresponding to different values of $n$ are orthogonal in terms of the inner product in $L_2(T)$. It can be verified by passing in the corresponding inner product to the coordinates $(r,\,s,\,\alpha)$, and then notice that the integral over the angle $\alpha$ will be equal to zero.

\subsection{Billiard corresponding to the quasimodes.}
\label{sec_bill}

It is known that semiclassical asymptotic functions are closely related to classical billiards \cite{Laz93, Laz88}. On the other hand, semiclassical asymptotics are also related to the Hamiltonian systems in the phase space \cite{BabBul72, Laz93, MF}. In that case, billiards can be constructed as projections of the trajectories of the Hamiltonian systems onto the configuration space. Such billiards, generally speaking, are not integrable, but the presence of different scales makes it possible to reduce the dimension of the problem  and obtain the integrable case up to small corrections. Similar  procedure was done in the works \cite{DobrMinShl18, DobrMinNeiSh19}, where the billiard was studied for the spectral problem of the stationary Schr\"{o}dinger equation in a narrow angle.

To the three-dimensional eigenproblem (\ref{Laplace_eig}) for the Laplacian, the three-dimensional Hamiltonian function 
$$
H_{3D}(x,\,p)=|p|^2=p_1^2+p_2^2+p_3^2
$$
corresponds. The projection onto the configurational space $x\in\mathbb{R}^3$ of the trajectories of the corresponding Hamilton system will give us the billiard mapping inside the solid torus $T$. The corresponding billiard trajectories will be piece-wise straight lines with reflection from the boundary $\partial T$ according with the Snell's law.

From the other point of view, we can make the change of variables (\ref{coord_sys_bound_3D}), (\ref{coord_sys_bound}) in the equation (\ref{Laplace_eig}) and pass to the two-dimensional equation (\ref{2D_eq_new}). Therefore, we can present the two-dimensional Hamiltonian 
\begin{gather}
\label{Hamiltonian_full}
H(s,\,p_s,\,\rho,\,p_\rho)=H_0(s,\,p_s)+h H_1(s,\,p_s,\,\rho,\,p_\rho),\\
\nonumber
H_0(s,\,p_s)=p_s^2+V(s;\,\mathcal{E}^2),\quad H_1(s,\,p_s,\,\rho,\,p_\rho)=p_\rho^2+\rho \mathcal{A}^3(s;\,\mathcal{E}^2).
\end{gather}
Because during the procedure for obtaining the equation (\ref{2D_eq_new}) we omitted terms of order $O(h^2)$, the Hamiltonain (\ref{Hamiltonian_full}) describes the properties of the dynamical system up to the correction $O(h^2)$. 

The corresponding system for the Hamiltonian (\ref{Hamiltonian_full}) has the following form
\begin{gather}
\label{2d_sys}
\dot{s}=2p_s,\quad \dot{p}_s=-V'(s;\,\mathcal{E}^2)-h \rho \frac{\partial \mathcal{A}^3(s;\,\mathcal{E}^2)}{\partial s},\quad
\dot{\rho}=h 2p_\rho,\quad \dot{p}_\rho=-h \mathcal{A}^3(s;\,\mathcal{E}^2).
\end{gather}

The projections of the trajectories of the system (\ref{2d_sys}) onto the configurational space $(s,\,\rho)$  give us the billiard, which has the caustic corresponding to the reflection from the potential. Using formula (\ref{w_main_sol}) we can describe this caustic, up to the $O(h^2)$, with the following equation
\begin{equation}
\label{caust}
\rho_c(s)=\frac{t_k}{\mathcal{A}(s;\,\mathcal{E})} \Leftrightarrow r_c(s)=\frac{h t_k}{\mathcal{A}(s;\,\mathcal{E})}.
\end{equation}
This curve is a curve where Airy function $Ai(-t_k+\rho \mathcal{A}(s;\,\mathcal{E}^2))$ equals to zero. In other words, we can describe curve (\ref{caust}) as a union of turning points of our dynamical system along the normal to the boundary. 



System (\ref{2d_sys}) is the system with proper degeneracy (see \cite{ArnlKozlNeist}, paragraph 5.2.1) and we can implement the averaging procedure in order to analyze this system. Because  (\ref{2d_sys}) is the system with reflection condition at the boundary, the averaging procedure has to take it into account, \cite{Laz88, DobrMinNeiSh19}.

 The unperturbed system corresponds to the one-dimensional Schr\"{o}dinger equation with the Hamiltonian $H_0(s,\,p_s)$ and its first integral will also be  an integral for the averaged system. Since the system (\ref{2d_sys}) corresponds to the two-dimensional case, we obtain after the averaging procedure the integrable system again. 


The unperturbed Hamilton system has the form
$$
\dot{s}=2p_s,\quad \dot{p}_s=-V'(s;\,E^2).
$$
Suppose that there are two turning points and let functions $(Q(\tau;\, E^2),\,P(\tau;\, E^2))$ denote solution to this system. These functions are periodic with respect to $\tau$ and let's $\mathcal{T}$ be a period.

It is known that the integral for this system is an action $I(E^2)$ and it satisfies  the following relation
$$
I(E^2)=\frac{1}{2\pi}\int\limits_{0}^{\mathcal{T}}P\dot{Q}d\tau=\varepsilon(m+\frac{1}{2}).
$$
This equality corresponds the Bohr-Sommerfeld quantization rule for the principle term of the spectral parameter, such that $\mathcal{E}^2=E^2+O(h)$. 

As we mentioned above, the averaged Hamilton system is integrable \cite{Laz88, DobrMinNeiSh19, ArnlKozlNeist}. We want to mention that because the averaging process reduces the Hamilton system to the integrable case (up to small correction $O(h^2)$ ), we also have possibility to proceed with operator separation of variables and reduce two-dimensional equation to the two one-dimensional ones (with the same small correction $O(h^2)$) similar to \cite{DobrMinNeiSh19}.

\section{Example.}
\label{sec_exampl}

Following \cite{OptWGM_wedge_1, OptWGM_wedge_2, Arnold}, as an example for illustration of the obtained results, we consider the construction of an asymptotic functions for the Laplace operator in the domain $T$ shown in Fig. \ref{pic2}. This region is obtained by rotating along the $z$-axis the region $\Omega$ bounded by a curve that can be described as a equilateral triangle with smoothed wedges  and it is shown in Fig. \ref{pic1}. 

Let us describe how we construct this curve. We fix the length of this curve $L=2\pi$, and set the curvature as follows
$$
k(s)=\frac{2\pi}{\gamma}f(s),\quad s\in[0,\,2\pi],\quad \gamma=\int\limits_{0}^{ 2\pi}f(\zeta)d\zeta,
$$
where is the function
$$
f(s)=e^{-s^2/\sigma^2}+e^{-(s-2\pi/3)^2/\sigma^2}+e^{-(s-4\pi/3)^2/\sigma^2}+e^{-(s-2\pi)^2/\sigma^2},\quad \sigma=0.4.
$$
We continue curvature $k(s)$ periodically over the interval $s\in[0,\,2\pi]$. Define the function
$$
a(s)=\int\limits_{0}^{s}k(\zeta)d\zeta-\gamma,
$$
then, following \cite{Pressey}, in the plane $(x,\,z)$ the equation for the curve $S$ has the form
$$
x=R+Q_1(s),\,z=Q_2(s),\quad Q_1(s)=\int\limits_{0}^{s}\cos(a(\zeta))d\zeta,\quad Q_2(s)=\int\limits_{0}^{s}\sin( a(\zeta))d\zeta.
$$

{\bf Remark 12.} If the area $\Omega$ is bounded by a circle (i.e., the three-dimensional area $T$ is bounded by a torus), then the functions $Q_{1,\,2}(s)$ are just trigonometrical functions.  Nevertheless, this form does not lead to significant simplifications for the asymptotic formulas (\ref{w_main_sol}).

We chose the following values for the parameters
$$
R=3,\quad h=0.015,\quad \varepsilon=h^{3/2}\approx 0.00183,\quad
n=1500 \Rightarrow \frac{n \varepsilon}{R}\approx 0.915.
$$
We also choose 
$$
m=5,\quad t_2\approx 4.0879.
$$
Here $m$ is the quantum number from the quantization rule (\ref{Bohr_Zommerfeld}) and $(-t_2)$ is the second root of the Airy function $Ai(-t_2)=0$.

For the given quantum number $m=5$ we have the turning points
$$
s_{-}\approx 3.7573,\quad s_{+}\approx 4.2266
$$
and the value for $E^2$ and the correction $E_1$
$$
E^2\approx 0.31169,\quad h t_2 E_1\approx 0.01587 \Rightarrow\quad  \mathcal{E}^2=E^2+ht_2 E_1\approx 0.327566.
$$

\begin{figure}[!h]
\begin{center}
\includegraphics[scale=.6]{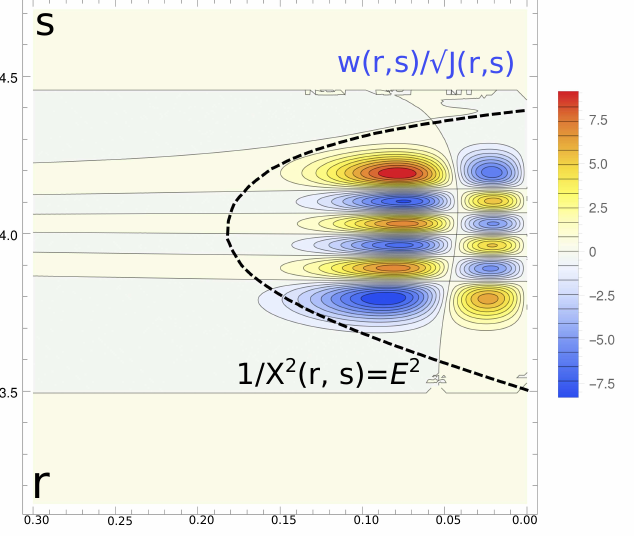}
\includegraphics[scale=.6]{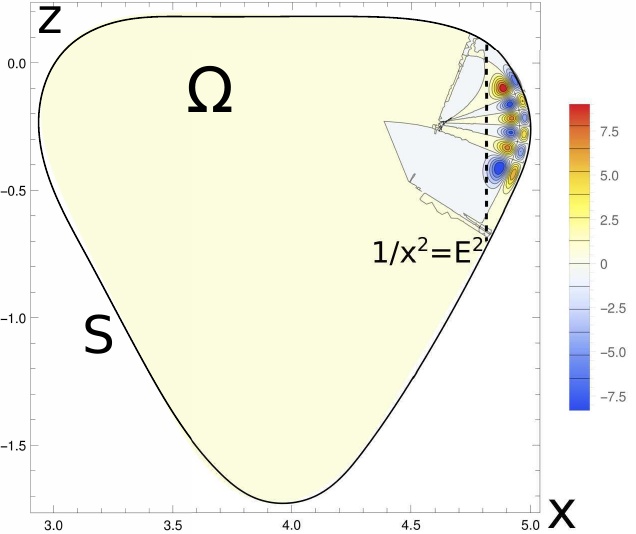}
\caption{On the left figure, the level lines of the function $\tilde{w}(r,\,s)/\sqrt{J(r,\,s)}$ and of the potential $1/X^2(r,\,s)=\mathcal{E}^2$ (bold black dashed line) with the given above parameters are shown. The area of the arguments $s\in[\pi,\,3\pi/2]$ and $r\in[0,\,0.3]$. The level lines of the same function are shown inside the area $\Omega$ with boundary $S$ in the plane $(x,\,z)$ on the right figure. The level line $1/x^2=\mathcal{E}^2$ of the potential is also shown as bold dashed line. \label{pic4}}
\end{center}
\end{figure}

The level lines of the function $\tilde{w}(r,\,s)/\sqrt{J(r,\,s)}$  are shown on fig. \ref{pic4} where function $\tilde{w}(r,\,s)$ is defined according to the (\ref{w_func_fin}), and $J(r, \,s)$ is the Jacobian of the transition from Cartesian to curvilinear coordinates. We demonstrated this function in two coordinate systems; on the left graph, we plotted the function in the rectangle $s\in[\pi,\,3\pi/2]$ and $r\in[0,\,0.3]$. On the right graph, we plotted this function inside the whole area $\Omega$ in coordinates $(x,\,z)$ and this function is localized near some part of the  boundary of the torus. Figure \ref{pic4} also shows the level line of the potential $1/X^2(r,\,s)=\mathcal{E}^2$, which arises in the two-dimensional equation (\ref{2D_eq}) and determines the asymptotic behavior in the Schrödinger equation.

\begin{figure}[!h]
\begin{center}
\includegraphics[scale=.7]{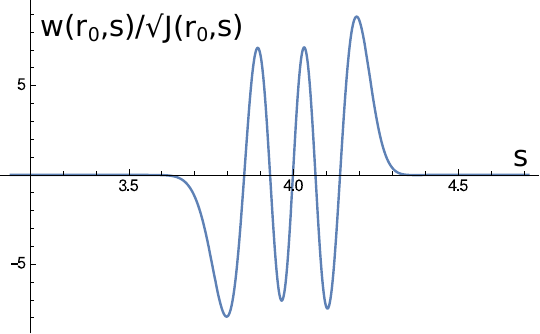}
\includegraphics[scale=.7]{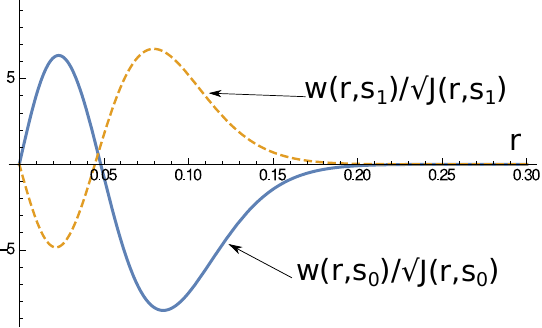}
\caption{The left figure shows the graph of the function $\tilde{w}(r_0,\,s)/\sqrt{J(r_0,\,s)}$ for $r_0=0.075$ and for $s\in[\pi,\, 3\pi/2]$. The right figure shows the graphs $\tilde{w}(r,\,s_i)/\sqrt{J(r,\,s_i)}$, $i=0,\,1$ where $s_0=3.8$ (solid line) and $s_1=3.9$ (dashed line). \label{pic7}}
\end{center}
\end{figure}

On fig. \ref{pic7}, we illustrate the form of the function $\tilde{w}(r,\,s)/\sqrt{J(r,\,s)}$ with one of the variables fixed. Namely, we illustrate the form of this function for $r_0=0.075$, as well as for different values of the variable $s$, namely, for the values $s_0=3.8$ and $s_1=3.9$.

\begin{figure}[h!]
\begin{center}
\includegraphics[scale=0.3]{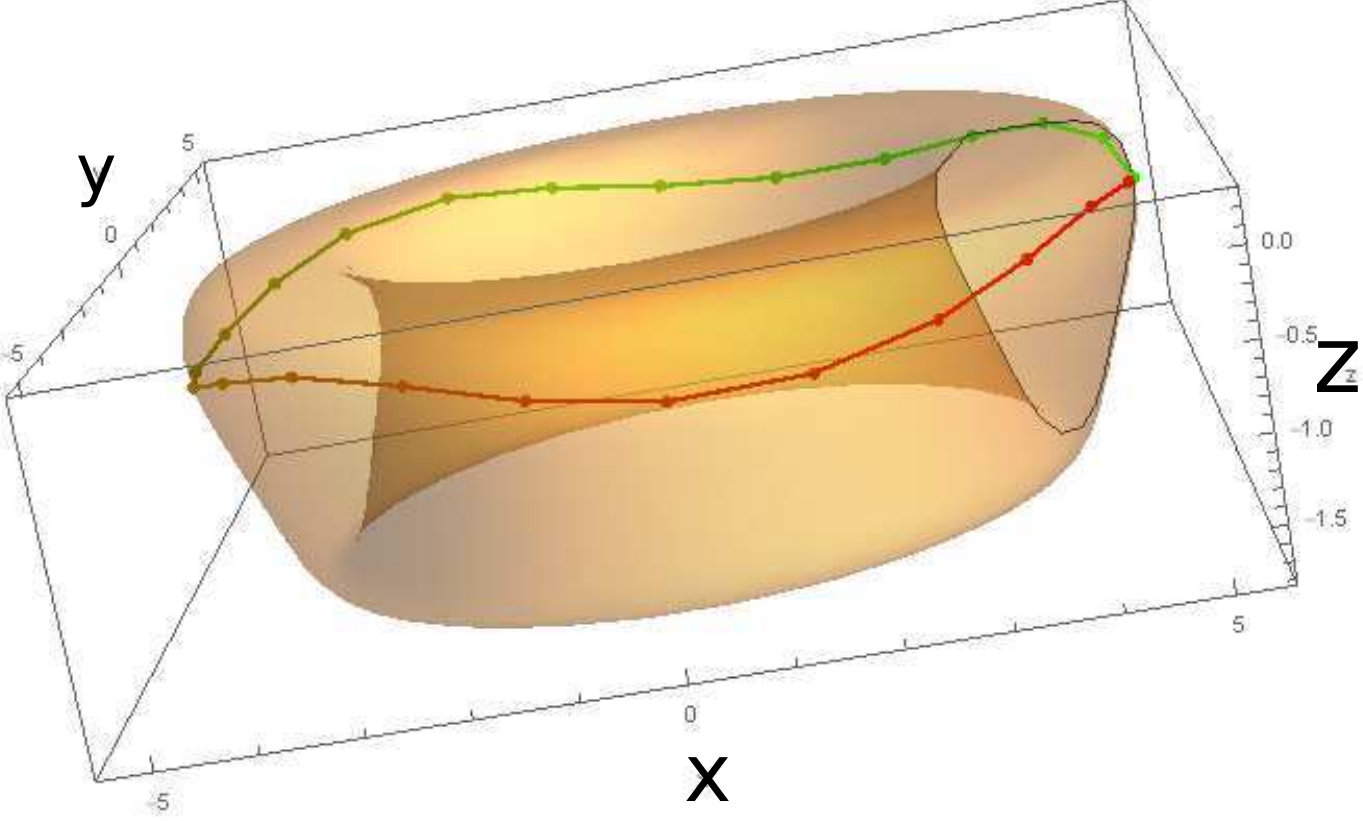}
\includegraphics[scale=1.15]{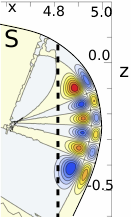}
\includegraphics[scale=.8]{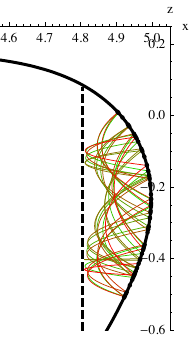}
\caption{The left figure shows two 3D billiard trajectories, points on them correspond to the reflection from the boundary. In the middle, we have drawn  the level lines of $\tilde{w}/\sqrt{J}$ (the dashed line is the level line of the potential). On the left picture, we drawn trajectories for the 2D billiard.  \label{pic8}}
\end{center}
\end{figure}

On the fig. \ref{pic8}, we illustrate the three-dimensional and two-dimensional billiards. We calculated two different billiard trajectories for the 3D case and also denoted the points on these trajectories  where they reflect from the boundary surface. Similar was done in the 2D case, we illustrated the corresponding  billiard. For the illustration of the localization of the asymptotics in the 2D case, we also showed function $\tilde{w}(r,\,s)/\sqrt{J(r,\,s)}$ inside the area of localization.

\subsection{Simplification of the 1D asymptotic for the equation (\ref{1D_eq_base}).}
\label{subsec_simpl}

In this section, we provide an asymptotic solution for the one-dimensional Schrödinger equation (\ref{1D_eq_base}) using parabolic cylinder functions \cite{Olver} uniformly for all values of the $s$ in the case of the small distance between truning points.  

For simplicity, we consider the situation with symmetrical about zero turning points for the one-dimensional Schr\"{o}dinger equaion (\ref{1D_eq_base}). In other words, we assume that turning points $s_{\pm}$  have the form $s_{\pm}=\pm s^*$ where $s^*>0$. We assume that the values of the parameter $\mathcal{E}^2$ are such that the distance between the turning points is small enough, namely,  $s^*=O(\sqrt{h})$.

We also assume that in a neighborhood of the point $s=0$ the following relations are true
\begin{equation}
\label{Q1_Q2_s0}
Q_1(s)=q_0-q_2 s^2+O(s^4),\quad q_0,\,q_2>0,\quad
Q_2(0)=0,\,Q_2'(0)>0.
\end{equation}

Under the above assumptions, the following expansions hold for the coefficients of the one-dimensional equation (\ref{1D_eq_base}):
\begin{gather}
\label{pot_expan}
\frac{a_n^2}{X^2(0,\, s)}\equiv\frac{a_n^2}{(R+Q_1(s))^2}=\frac{a_n^2}{( R+q_0)^2}+s^2 \beta+O(s^4),\quad \beta=\frac{2q_2 a_n^2}{(R+q_0)^3}, \\
\label{A_expan}
\mathcal{A}^2(s;\, E^2)=\mathcal{A}^2(0,\; E^2)+s \mathcal{A}_1(0;\, E^2) +O(s^2).
\end{gather}

\begin{statement}
\label{Herm_pol}
Let us fix positive integer numbers $n,\,k,\,m$ and let the spectral parameter has the form 
$$
\mathcal{E}_{n,k,m}^2=E_{n,k}^2+ht_k E_1,
$$
where
\begin{equation}
\label{spec_pol}
E_{n,m}^2=\frac{a_n^2}{(R+q_0)^2}+\varepsilon(m+\frac{1}{2})2\sqrt{\beta},\quad
E_1=\mathcal{A}^2(0;\, E_{n, m}^2).
\end{equation}

Let $s^*=O(\sqrt{h})$ and let us define function
\begin{equation}
\label{Parabol_sol}
\psi(s)=A_0 D_{m}\left(\frac{s}{\sqrt{\varepsilon}}\sqrt{2}\beta^{1/4}+\frac{h}{\sqrt{2}\sqrt{\varepsilon}}t_k \frac{\mathcal{A }_1(0;\, E^2)}{\beta^{3/4}}\right),
\end{equation}
where constant $A_0>0$ is chosen in a way that $\|\psi(s)\|_{L_2[0,\,L]}=1$ and $D_m(\eta)$ is the parabolic cylinder function.

Then the function (\ref{Parabol_sol}) satisfies the relation $\|\hat{L}_0\psi(s)\|=O(h^2)$, where $\hat{L}_0$ is the one-dimensional operator defined in (\ref{L_oper}).

\end{statement}

{\bf Proof of the statement.} Since $|s|\le s^*=O(\sqrt{h})$ we can  simplify the equation (\ref{1D_eq_base}). In order to do so, we substitute the expansions (\ref{pot_expan}) and (\ref{A_expan}) into this equation. After simplification, we obtain the equation $\tilde{L}_0\psi(s)=O(h^2)$ with operator
\begin{equation}
\label{1D_eq_small}
\tilde{L}_0\equiv \hat{p}_s^2+v(E^2,\,E_1)+s^2 \beta+h t_k s\mathcal{A}_1(0;\, E^2),
\end{equation}
where we denoted
$$
v(E^2,\,E_1)=\frac{a_n^2}{(R+q_0)^2}-E^2+h t_k (\mathcal{A}^2(0;\, E^2 )-E_1),\quad
$$

The exact solution to the equation $\tilde{L}_0\psi(s)=0$ is the function $D_\nu(\eta(s))$, where argument $\eta(s)$ is the same argument as in (\ref{Parabol_sol}). The order $\nu$ of the function $D_{\nu}(\eta(s))$ has complicated form, and therefore we do not provide it here. From the other hand,  in our case, the quantization rule is $\nu$ has to be a positive integer $m$. After simplification of the order $\nu$ we obtain the following quantization rule
$$
\frac{1}{2\sqrt{\beta}}\left(E^2-\frac{a_n^2}{(R+q_0)^2}-ht_k(\mathcal{A}^2(0; \, E^2)-E_1)\right)=\varepsilon(m+\frac{1}{2})+O(h^2).
$$
From this relation, we  obtain (\ref{spec_pol}). The estimate for the norm is the consequence of \cite{ADNTs}. {\bf Statement \ref{Herm_pol} is proven.}

\section{Operator separation of variables. Proof of the statement \ref{st_one_dim_eq}.}
\label{sec_sep_var}

In this section, we provide a proof for the  statement \ref{st_one_dim_eq}. This statement is the key assertion for the proof of the main results and the proof of it based on the operator separation of variables  \cite{GrDobrSergT16, BrGrDobr12}.

Let us recall the operator $\hat{H}$ in the equation (\ref{2D_eq_new})
$$
\hat{H}=H_0(s,\,\hat{p}_s;\, \mathcal{E}^2)+h\hat{D},\quad
H_0=\hat{p}_s^2+V(s;\, \mathcal{E}^2),\quad
\hat{D}=-\frac{\partial^2}{\partial \rho^2}+\rho \mathcal{A}^3(s;\, \mathcal{E}^2).
$$
Potential $V(s;\,\mathcal{E}^2)$ and function $\mathcal{A}(s;\,\mathcal{E}^2)$ are defined in (\ref{potential_A_coef}).

Equation (\ref{2D_eq_new}) has the form $\hat{H}w=0$ and we are looking for a solution to this equation (\ref{2D_eq_new}) in the following form
$$
w(\rho,\,s)=\chi(\rho,\,\stackrel{2}{s},\,\stackrel{1}{\hat{p}}_s;\, \mathcal{E}^2;\,\varepsilon,\,h)\psi(s),\quad L(\stackrel{2}{s},\,\stackrel{1}{\hat{p}}_s;\, \mathcal{E}^2\,;\varepsilon,\,h)\psi(s)=0,\quad \chi(0,\,\stackrel{2}{s},\,\stackrel{1}{\hat{p}}_s;\, \mathcal{E}^2;\,\varepsilon,\,h)=0.
$$
Here the operators $\hat{\chi}$ and $\hat{L}$ are to be defined, and $\mathcal{E}^2$ is the spectral parameter. The last equality is the Dirichlet boundary conditions (\ref{chi_boundary}).


Let us now substitute the function $w(\rho,\,s)$ in the given form into the equation (\ref{2D_eq_new}). This substitution leads to the equation $H\hat{\chi}\psi(s)=0$, and if the function $\psi(s)$ is a solution to the one-dimensional equation $\hat{L}\psi(s)=0$, then the function $w(\rho,\,s)$ will be a solution to the equation (\ref{2D_eq_new}) if the operator equality (\ref{operator_equal}) holds:
$$
\hat{H}\hat{\chi}=\hat{\chi}\hat{L}.
$$

Let us pass in this equality  to the symbols for the operators $\hat{H}\hat{\chi}$ and $\hat{\chi}\hat{L}$ as operators with respect to variable $s$. This leads us to the differential problem with respect to the $\rho$  for the so-called operator-valued symbols \cite{GrDobrSergT16, Serg22, BrGrDobr12}. 

For symbols of operators $\hat{\chi}$ and $\hat{L}$, we assume that the following expansions are valid
\begin{gather*}
\chi(\rho,\,s,\,p_s;\, \mathcal{E}^2;\varepsilon,\,h)=\chi_0(\rho,\,s,\,p_s;\, \mathcal{E}^2;\, h)-i\varepsilon \chi_1(\rho, \,s,\,p_s;\, \mathcal{E}^2;\, h)+O(\varepsilon^2),\\
L(s,\,p_s;\, \mathcal{E}^2;\varepsilon,\,h)=L_0(,s,\,p_s;\, \mathcal{E}^2;\, h)-i\varepsilon L_1(s,\,p_s;\, \mathcal{E}^2;\, h)+O(\varepsilon ^2).
\end{gather*}
Writing down the symbols for $\hat{H}\hat{\chi}$ and $\hat{\chi}\hat{L}$ and substitution of these symbols into the operator equality (\ref{operator_equal}) gives us the following equality 
\begin{equation}
\label{symb_equal}
\Bigl(H_0+h\hat{D}\Bigr)\chi_0-i\varepsilon\Bigl(H_{0p_s}\chi_{0s}+(H_0+h\hat{D})\chi_1\Bigr)=\chi_0L_0-i\varepsilon(\chi_0L_1+\chi_{0p_s}L_{0s}+\chi_1L_0)+O(\varepsilon^2).
\end{equation}
This equality leads to the equalities (\ref{symb_equal_L0}) and (\ref{symb_equal_L1}) after taking the coefficients with similar powers of parameter $\varepsilon$.

Assume that the functions $\chi_{0,\,1}$ and $L_{0,\,1}$ can be represented by the following series, generally speaking asymptotic, in powers of $\sqrt{h}$
\begin{equation}
\label{chi_L_expan}
\chi_0=\sum\limits_{k=0}^{2}h^{k/2}\chi_0^k,\quad \chi_1=\sum\limits_{k=0}^{+\infty} h^{k/2}\chi_1^k,\quad
L_0=\sum\limits_{k=0}^{2}h^{k/2}L_0^k,\quad
L_1=\sum\limits_{k=0}^{+\infty}h^{k/2}L_1^k.
\end{equation}
The terms of these expansions $\chi_i^k$ depend on the variables $(\rho,\,s,\,p_s)$, and the functions $L_i^k$ depend on $(s,\,p_s)$. For brevity, we did not specify this dependence in (\ref{chi_L_expan}). 
Equality (\ref{symb_equal}) together with (\ref{chi_L_expan}) allows us to calculate the symbols $\chi$ and $L$ with the given precision
 as a series in terms of powers $\varepsilon^{1/3}=h^{1/2}$ by analogy with \cite{BabBul72, Popov20}.

Before the calculation of the elements in the expansions (\ref{chi_L_expan}) we need to rewrite the normalization condition (\ref{normal_chi}) for operator $\hat{\chi}$ in terms of the expansion  (\ref{chi_L_expan}).
\begin{lemma}
\label{lm_norm_cond_coef}
Let operator $\hat{\chi}$ satisfy the normalization condition (\ref{normal_chi}) and let the expansions (\ref{chi_L_expan}) be valid for that operator. Then we have the following conditions 
\begin{equation}
\label{norm_cond_h}
\int\limits_{0}^{+\infty}|\chi_0^0|^2d\rho=1,\quad \int\limits_{0}^{+\infty}(\chi_0^0\overline{\chi}_0^1+\chi_0^1\overline{\chi}_0^0)d\rho=0,\quad 
\int\limits_{0}^{+\infty}(|\chi_0^1|^2+\chi_0^0\overline{\chi}_0^2+\chi_0^2\overline{\chi}_0^0)d\rho=0,
\end{equation}
where the bar stands for the complex conjugation.
\end{lemma}

{\bf Proof.} The adjoint operator $\hat{\chi}^*$ can be written in the following form
$$
\hat{\chi}^*=\overline{\chi}(\stackrel{2}{\hat{p}_s},\,\stackrel{1}{s},\,\rho;\, h,\,\varepsilon),
$$
where line stands for the complex conjugation. 
We need to rearrange the action of the operators $\hat{p}_s$ and $s$ in this operator since we work with inverse ordering. With the help of \cite{StShN} we obtain  the following formula 
$$
smb(\hat{\chi}^*)=\overline{\chi}_0(p_s,\,s,\,\rho;\, h)-i\varepsilon\left(\frac{\partial^2\overline{\chi}_{0}}{\partial p_s\partial s}(p_s,\,s,\,\rho;\, h)+\overline{\chi}_1(p_s,\,s,\,\rho;\, h)\right)+O(\varepsilon^2).
$$
for the corresponding symbol. Here we also used the expansion of the symbol $smb(\hat{\chi})$ with respect to the $\varepsilon$.
After we get this expansion, we can present the symbol $smb(\hat{\chi}^*\hat{\chi})$ in the following form 
$$
smb(\hat{\chi}^*\hat{\chi})=|\chi_0|^2-i\varepsilon \left(\left(\frac{\partial^2 \overline{\chi}_{0}}{\partial p_s\partial s}+\overline{\chi}_1\right)\chi_0+\overline{\chi}_0\chi_1+\frac{\partial \overline{\chi}_0}{\partial p_s}\frac{\partial \chi_0}{\partial s}
\right)+O(\varepsilon^2).
$$
This gives us the following conditions
$$
\int\limits_{0}^{+\infty}|\chi_0|^2d\rho=1,\quad \int\limits_{0}^{+\infty}\left(\frac{\partial}{\partial s}(\overline{\chi}_{0p_s}\chi_0)+\overline{\chi}_1\chi_0+\overline{\chi}_0\chi_1\right)d\rho=0.
$$
After substitution of the expansion (\ref{chi_L_expan}) for $\chi_0$ into the first integral equality, we obtain the statement of the lemma. {\bf Lemma \ref{lm_norm_cond_coef} is proven.}

\begin{lemma}
\label{lm_oper_1}
Under the above assumptions, for the terms of the expansion (\ref{chi_L_expan}) the following equalities hold:
\begin{gather*}
L_0^0(s,\,p_s)=H_0\equiv p_s^2+\frac{a_n^2}{X^2(0,\,s)}-\mathcal{E} ^2,\quad L_0^1\equiv 0,\quad L_0^2=t_k \mathcal{A}^2(s;\, \mathcal{E}^2),\\
\chi_0^0=\frac{\sqrt{\mathcal{A}(s;\, \mathcal{E}^2)}}{|Ai'(-t_k)|} Ai(-t_k+\rho \mathcal{A}(s;\, \mathcal{E}^2)).
\end{gather*}
Here the number $(-t_k)$ is the $k$-th root of the Airy function: $Ai(-t_k)=0$.
\end{lemma}

{\bf Proof of the lemma. } Let us substitute the expansions (\ref{chi_L_expan}) into the equality (\ref{symb_equal}) and group the coefficients for $h^0=1$, $h^{1/2}$ and $h$. This procedure leads to the following equality
$$
H_0(\chi_0^0+\sqrt{h}\chi_0^1+h\chi_0^2)+h\hat{D}\chi_0^0=L_0^0(\chi_0^0 +\sqrt{h}\chi_0^1+h \chi_0^2)+\sqrt{h}L_0^1(\chi_0^0+\sqrt{h}\chi_0^1)+h L_0^2\chi_0^ 0.
$$
For the main term, we have the equality $H_0\chi_0^0=\chi_0^0L_0^0$. Since $H_0$ is not a differential operator in $\rho$, and the function $\chi_0^0$ depends on $\rho$, we immediately obtain the expression for $L_0^0$. 

Further, after cancellation of terms $H_0\chi_0^1$ and $\chi_0^1L_0^0$, since there are no terms of order $\sqrt{h}$ on the left side of the equality, we also get that $L_0^1\equiv 0$.

Finally, the equation for $\chi_0^0$ and $L_0^2$ remains
\begin{equation}
\label{eigen_chi0}
\hat{D}\chi_0^0=L_0^2\chi_0^0.
\end{equation}
We regard this equation as spectral with respect to the operator $\hat{D}$, where $\chi_0^0$ plays the role of an eigenfunction and $L_0^2$ plays the role of an eigenvalue.
The solution to the equation (\ref{eigen_chi0}) is the Airy function
$$
\chi_0^0(\rho,\,s)=Ai\left(-\frac{L_0^2}{\mathcal{A}^2(s;\, \mathcal{E}^2)}+\rho \mathcal{A}(s;\, \mathcal{E}^2)\right).
$$
The boundary condition $\chi_0^0|_{\rho=0}=0$ leads to the formula for $L_0^2$. 

All that is left is to use the normalization condition (\ref{norm_cond_h}). We have
$$
\int\limits_{0}^{+\infty}|\chi_0^0|^2d\rho=\|\chi_0^0\|^2=\frac{1}{\mathcal{A}(s;\, \mathcal{E}^2)}\left(Ai'(-t_k)\right)^2,
$$
which gives the formula for $\chi_0^0$. {\bf Lemma \ref{lm_oper_1} is proven. }

The next terms in the expansions (\ref{chi_L_expan}) are defined sequentially. The functions $\chi_{0,\,1}^k$ are determined from an inhomogeneous equation of the form $(\hat{D}-L_0^2)\chi_{0,\,1}^k=F_{0,\,1}^k$, where $F_{0,\,1}^k$ is some function that includes unknown terms $L_{0,\,1}^k$ from the expansions (\ref{chi_L_expan}). The condition for the solvability of this inhomogeneous equation is the orthogonality of the function $F_i^k$ to the solution of the homogeneous equation (\ref{eigen_chi0}), i.e., functions $\chi_0^0$. This condition leads to the equation for the unknown functions $L_{0,\,1}^k$.

The next lemma demonstrates this approach for determining $\chi_0^1$, as well as  $L_1^0$.

\begin{lemma}
\label{lm_oper_2}
The following formulas are valid
\begin{equation}
\label{chi01_Ai}
L_1^0\equiv 0,\quad 
\chi_0^1=-i\frac{1}{Ai'(-t_k)}\frac{\mathcal{A}'(s;\, \mathcal{E}^2)}{2\sqrt{\mathcal{A}(s; \, \mathcal{E}^2)}}\rho^2p_s Ai(-t_k+\rho \mathcal{A}(s;\, \mathcal{E}^2)).
\end{equation}
\end{lemma}

{\bf Proof of the lemma. } Let us combine coefficients with the degree $h^{3/2}=\varepsilon$ in (\ref{symb_equal}) taking into account (\ref{chi_L_expan}) and the results of Lemma \ref{lm_oper_1}. We arrive at the following equation
$$
(\hat{D}-t_k\mathcal{A}^2(s;\, \mathcal{E}^2))\chi_0^1=i(2p_s \chi_{0s} ^0-L_1^0\chi_0^0).
$$
Here we substituted  $\partial H^0_{0}/\partial p_s=2p_s$.

The solvability condition  for this inhomogeneous equation (orthogonality of the right-hand side of the equation to $\chi_0^0$) leads to the following equality, from which we determine $L_1^0$
$$
i(2p_s (\chi_{0s}^0,\, \chi_0^0)-L_1^0 )=0.
$$
Here we used $(\chi_0^0,\,\chi_0^0)=1$. Because of this normalization we have
$$
\frac{d}{ds}(\chi_0^0,\,\chi_0^0)=2(\chi_{0s}^0,\,\chi_0^0)=0,
$$
and thus $L_1^0\equiv 0$.

As a result, we get the following equation for $\chi_0^1$ 
$$
(\hat{D}-t_k\mathcal{A}^2(s;\, \mathcal{E}^2))\chi_0^1=i2p_s \chi_{0s}^0.
$$
By direct substitution of the function (\ref{chi01_Ai}) into this equation, one can verify that it is a solution. 

Since function $\chi_0^0$ is real and function $\chi_0^1$ is purely imaginary, the normalization condition for $\chi_0^1$ in (\ref{norm_cond_h}) is also satisfied. {\bf Lemma \ref{lm_oper_2} is proven. }

The  lemmas \ref{lm_oper_1} and \ref{lm_oper_2}  lead us to the formulas (\ref{L_oper}) and (\ref{chi0}) for symbols of $L_0$ and $\mathcal{X}_0$.

The next lemma leads to the inequality (\ref{oper_eq_real}).

\begin{lemma}
\label{lm_oper_R}
Operator $\hat{R}\equiv \hat{H}\hat{\mathcal{X}}_0-\hat{\mathcal{X}}_0\hat{L}_0$ has the following form
$$
\hat{R}=-ih^2\Bigl(\hat{\chi}_{0s}^1\hat{p}_s+\hat{\chi}_{0p_s}^1 V'(s;\,\mathcal{E}^2)\Bigr)-h^3\Bigl(\chi_{0ss}^0+t_k\hat{\chi}_{0p_s}^1 (\mathcal{A}^2)_s\Bigr)-h^2\varepsilon\hat{\chi}_{0ss}^{1}+h\bigl(2\chi_{0\rho}^0\partial_\rho+\chi_0^0\partial_\rho^2\bigr)+h^{3/2}\bigl(2\hat{\chi}_{0\rho}^1\partial_\rho+\hat{\chi}_0^1\partial_\rho^2\bigr).
$$
\end{lemma}

{\bf Proof.} Let us calculate the operators $\hat{H}\hat{\chi}_0$ and $\hat{\chi}_0\hat{L}_0$  taking into account that in these operators operator of differentiation acts first. With the help of lemmas \ref{lm_oper_1} and \ref{lm_oper_2}, we get the following equalities
\begin{gather*}
\hat{H}\hat{\chi}_0=\hat{H}_0\chi_0^0+ht_k\mathcal{A}^2\chi_0^0+\sqrt{h}\hat{H}_0\hat{\chi}_0^1+h^{3/2}\hat{D}\hat{\chi}_0^1+h\bigl(2\chi_{0\rho}^0\partial_\rho+\chi_0^0\partial_\rho^2\bigr)+h^{3/2}\bigl(2\hat{\chi}_{0\rho}^1\partial_\rho+\hat{\chi}_0^1\partial_\rho^2\bigr),\\
\hat{\chi}_0\hat{L}_0=\hat{H}_0\chi_0^0-[\hat{H}_0,\,\chi_0^0]+ht_k\mathcal{A}^2\chi_0^0+\sqrt{h}\hat{H}_0\hat{\chi}_0^1-\sqrt{h}[\hat{H}_0,\,\hat{\chi}_0^1]+h^{3/2}t_k\mathcal{A}^2\hat{\chi}_0^1-h^{3/2}[t_k\mathcal{A}^2,\,\hat{\chi}_0^1],
\end{gather*}
where $[\hat{A},\,\hat{B}]$ is the commutator of two operators $\hat{A}$ and $\hat{B}$. The corrections in the brackets appear after calculation of the operator $\partial_\rho^2(\hat{\chi}_0^0+\sqrt{h}\hat{\chi}_0^1)$.
Therefore, we have the equality
$$
\hat{R}=[\hat{H}_0,\,\chi_0^0]+\sqrt{h}[\hat{H}_0,\,\hat{\chi}_0^1]+h^{3/2}(\hat{D}-t_k\mathcal{A}^2)\hat{\chi}_0^1+h^{3/2}[t_k\mathcal{A}^2,\,\hat{\chi}_0^1]+h\bigl(2\chi_{0\rho}^0\partial_\rho+\chi_0^0\partial_\rho^2\bigr)+h^{3/2}\bigl(2\hat{\chi}_{0\rho}^1\partial_\rho+\hat{\chi}_0^1\partial_\rho^2\bigr).
$$
Further calculation gives the following
\begin{gather*}
h^{3/2}(\hat{D}-t_k\mathcal{A}^2)\hat{\chi}_0^1=h^{3/2}i 2\chi_{0s}^0\hat{p}_s,\quad
[\hat{H}_0,\,\chi_0^0]=[\hat{p}_s^2,\,\chi_0^0]=-i\varepsilon 2\chi_0^0\hat{p}_s-\varepsilon^2\chi_{0ss}^0,\\
[\hat{H}_0,\,\hat{\chi}_0^1]=-i\varepsilon 2\hat{\chi}_{0s}^1\hat{p}_s-i\varepsilon\hat{\chi}_{0p_s}^1 V'(s;\,\mathcal{E}^2) -\varepsilon^2\hat{\chi}_{0ss}^1,\quad
[t_k\mathcal{A}^2,\,\hat{\chi}_0^1]=-\varepsilon t_k\hat{\chi}_{0p_s}^1 (\mathcal{A}^2)_s.
\end{gather*}
After substitution into the definition of the operator $\hat{R}$, we will obtain the statement of the lemma. {\bf Lemma \ref{lm_oper_R} is proven.}

After this lemma the estimate (\ref{oper_eq_real}) on the symbol of the operator $\hat{R}$ becomes trivial.
This completes the proof of the statement \ref{st_one_dim_eq}.

\section{Derivation of the two-dimensional equation.}
\label{sec_2D_eq_vyvod}
Recall that the solid torus $T$ is obtained by rotating the domain $\Omega$ bounded by the convex curve $S$. In this section, we give a derivation of the two-dimensional equations (\ref{2D_eq}) and (\ref{2D_eq_new}) in curvilinear coordinates $(r,\,s)$ in a neighborhood of the curve $S$. 

Let the curve $S$ be described in the plane $(x,\,z)$ by the equations
$$
x=R+Q_1(s),\quad z=Q_2(s),
$$
where $s$ is the arclength parameter of the curve, and the functions $Q_{1,\,2}(s)$ are periodic with period $L$, where $L$ is the length of the curve $S$. Let the coordinate $r$ be directed along the inner normal to $S$. 
Then in the neighborhood of the surface $\partial T$ we have a coordinate system $(r,\,s,\,\alpha)$ defined in (\ref{coord_sys_bound_3D}), (\ref{coord_sys_bound})
\begin{gather*}
x(r,\,s,\,\alpha)=X(r,\,s)\sin\alpha,\quad y(r,\,s,\,\alpha)=X(r,\,s )\cos\alpha,\quad z(r,\,s,\,\alpha)=Z(r,\,s),\\
X(r,\,s)=(R+Q_1(s))-r Q_2'(s),\quad Z(r,\,s)=Q_2(s)+r Q_1'(s).
\end{gather*}

The Lam\'{e} coefficients of the transition to the new coordinate system $(r,\,s,\,\alpha)$ have the form
$$
h_r=1,\quad h_\alpha=X(r,\,s),\quad h_s=(1-rk(s)),
$$
where $k(s)$ is the curvature of the curve $S$.

The coefficients $h_r$ and $h_\alpha$ are trivially calculated. Let's calculate coefficient $h_s$:
$$
h_s^2=Q_1^{\prime 2}-2r Q_1'Q_2''+r^2 Q_2^{\prime\prime 2}+Q_2^{\prime 2}+2r Q_2'Q_1''+r^2 Q_1^{\prime\prime 2}.
$$
Since $Q_1(s)$ and $Q_2(s)$ describe the curve $S$ in terms of an arclength parameter $s$, the vector $(Q_1'(s),\,Q_2'(s))$ is a unit vector tangent to $S$. The vector $(Q_1''(s),\,Q_2''(s))$ will be collinear to the normal vector, and according to the Frenet-Serret  formulas \cite{Pressey}, its length is equal to the curvature $k(s)$. Further, the normal vector has the form $(-Q_2'(s),\,Q_1'(s))$, so the expression $Q_1'Q_2''-Q_2'Q_1''$ is the inner product of the normal vector and the vector collinear to the normal, so this expression  equals to the length of the normal vector, i.e. curvature $k(s)$. Summarazing all the above, we arrive at the following equality
$$
h_s^2=1-2r k(s)+r^2 k^2(s)=(1-rk(s))^2.
$$

The Jacobian of the transition to the coordinates $(r,\,s,\,\alpha)$ equals to
$$
J=h_r h_s h_\alpha=(1-rk(s))X(r,\,s).
$$

Thus the equation $-\Delta u=\lambda^2 u$ in the coordinates $(r,\,s,\,\alpha)$ takes the form
$$
-\frac{1}{J}\left(\frac{\partial}{\partial r}J\frac{\partial u}{\partial r}+\frac{\partial}{\partial s}\frac {J}{(1-rk(s))^2}\frac{\partial u}{\partial s}+\frac{(1-rk(s))^2}{J}\frac{\partial ^2 u}{\partial \alpha^2}\right)=\lambda^2 u.
$$

Separating the angle $\alpha$ in the function $u(r,\,s,\,\alpha)$ and separating the Jacobian $J$, we pass to the new function $w(r,\,s)$ as follows
$$
u(r,\,s,\,\alpha)=\frac{w(r,\,s)}{\sqrt{|J|}}e^{in \alpha}.
$$
After direct substitution of this form into the equation, we obtain the following equation for the function $w$
\begin{equation}
\label{2D_eq_vspom}
-\left(\varepsilon^2\frac{\partial^2 w}{\partial r^2}+\varepsilon^2\frac{1}{(1-rk(s))^2}\frac{\partial^2 w}{\partial s^2}-\frac{\varepsilon^2 n^2}{X^2(r,\,s)}w-\varepsilon^2\frac{2 rk'(s )}{(1-rk(s))^3}\frac{\partial w}{\partial s}+O(\varepsilon^2)\right)=\mathcal{E}^2 w.
\end{equation}
Here we used the equality $\lambda=\mathcal{E}/\varepsilon$ and $O(\varepsilon^2)$ is a smooth and bounded function of order $\varepsilon^2$, which does not contain derivatives of the function $w$. In order to obtain the equation (\ref{2D_eq}), we just need to pass in this equation  to the operator $\hat{p}_s=i\varepsilon\partial/\partial s$ and write down it in the divergent form.

Let us now demonstrate how equation (\ref{2D_eq_new}) can be derived from the equation (\ref{2D_eq}) or from (\ref{2D_eq_vspom}).
We multiply equation (\ref{2D_eq_vspom}) on $(1-h\rho k(s))^2$ and after that move every term of the equation into the left-hand side. Thus, we obtained equation 
\begin{gather*}
(1-h\rho k(s))^2(\Delta_2-\mathcal{E}^2)= \hat{p}_s^2+a_n^2\frac{(1-h\rho k(s))^2}{X^2(h \rho,\,s)}-(1 -h \rho k(s))^2\mathcal{E}^2+\\
+h(1-h \rho k(s))^2\left(-\frac{\partial^2}{\partial \rho ^2}\right)-i\varepsilon h \frac{2 \rho k'(s)}{(1-h \rho k(s))}\hat{p}_s+O(\varepsilon^2)=0.
\end{gather*}
We expand coefficients in this equation with the precision $O(h^2)$, and it leads to the equation (\ref{2D_eq_new}). Note,  we have the following Taylor expansion with respect to $h$
$$
a_n^2\frac{(1-h\rho k(s))^2}{X^2(h\rho,\,s)}=\frac{a_n^2}{X^2(0,\,s)}+h\rho \left(\frac{2 a_n^2}{X^3(0,\,s)}Q_2'(s)-2 k(s)\frac{a_n^2}{X^2(0,\,s)}\right)+O(h^2).
$$


\begin{thebibliography}{99}
\bibitem{Airy} Sir George Biddell Airy, On sound and atmospheric vibrations with the mathematical elements of music. // London Cambridge, Macmillan, 1871

\bibitem{Rayleigh} Lord Rayleigh, The problem of the whispering gallery. // Philos. Mag., vol. 20, pp. 1001-1004, 1910, https://doi.org/10.1080/14786441008636993

\bibitem{Raman} C.V. Raman, G.A. Sutherland, On the whispering-gallery phenomenon. // Proceedings of the Royal Society A: Mathematical, Physical and Engineering Sciences, 1922,  vol. 100, issue 705, pp. 424-428, https://doi.org/10.1098/rspa.1922.0007

\bibitem{KatsPetr19} B.G. Katsnelson, P.S. Petrov, Whispering gallery waves localized near circular isobaths in shallow water. // JASA, 2019, vol. 146, pp. 1965-1978, https://doi.org/10.1121/1.5125419

\bibitem{PetrAnt20} P.S. Petrov, X. Antoine, Pseudodifferential adiabatic mode parabolic equations in curvilinear coordinates and their numerical solutions. // J. Of Computational Physics,  vol. 410,  2020, 109392, https://doi.org/10.1016/j.jcp.2020.109392

\bibitem{OptWGM1}  M. Sumetsky, Lasing microbottles. // Light Sci Appl, 2017, vol. 6, e17102, DOI: https://doi.org/10.1038/lsa.2017.102

\bibitem{OptWGM2} M. Foreman, J. Swaim, and F. Vollmer, Whispering gallery mode sensors. // Adv. Opt. Photon., 2015, vol. 7, pp. 168-240, DOI: https://doi.org/10.1364/AOP.7.000168

\bibitem{OptWGM_Arxiv} S. Suebka, E. McLeod, and J. Su, Ultra-high-Q free space coupling to microtoroid resonators. // 2023,	arXiv:2308.00726v1 [physics.optics], https://doi.org/10.48550/arXiv.2308.00726

\bibitem{OptWGM_wedge_1} H. Lee, T. Chen, J. Li, et al., Chemically etched ultrahigh-Q wedge-resonator on a silicon chip. // Nature Photonics, 2012,  vol. 6, pp. 369-373, DOI: https://doi.org/10.1038/nphoton.2012.109

\bibitem{OptWGM_wedge_2} T.J. Kippenberg, J. Kalkman, A. Polman, and K.J. Vahala, Demonstration of an erbium-doped microdisk laser on a silicon chip. // Phys. Rev. A, 2006, vol. 74, issue 5, 051802, DOI: 10.1103/PhysRevA.74.051802

\bibitem{KellerRub60} J.B. Keller, S.I. Rubinow, Asymptotic Solution of Eigenvalue Problems. // Annals of Physics, 1960, vol. 9, pp. 24-75, https://doi.org/10.1016/0003-4916(60)90061-0

\bibitem{BabBul72} V.M. Babich, V.S. Buldyrev. Asymptotic Methods in Short-wavelength Diffraction Theory. // Springer, 2011, Softcover reprint of the original 1st ed. 1972 edition


\bibitem{Kirpich79} N.Ya. Kirpichnikova. Uniform Asymptotics of Eigenfunctions of Whispering Gallery Type // J. Math Sci 19, 1366–1372 (1982). https://doi.org/10.1007/BF01085026

\bibitem{Laz67} V.F. Lazutkin, The asymptotics of the eigenfunctions of the Laplace operator, concentrated near the boundary of a region, // USSR Computational Mathematics and Mathematical Physics, vol. 7, issue 6, 1967, pp. 37-52, https://doi.org/10.1016/0041-5553(67)90115-2.

\bibitem{Laz93} V.F. Lazutkin, KAM Theory and Semiclassical Approximations to Eigenfunctions. //Springer-Verlag, 1993

\bibitem{Laz88} V.F. Lazutkin, Semiclassical asymptotics eigenfunctions. // Partial Differential Equations V. Asymptotic Methods for Partial Differential Equations.  Encyclopaedia of Mathematical Sciences (EMS, volume 34), Springer, 1999.

\bibitem{Arnold} V.I. Arnold, Modes and quasimodes. // Func. Anal. Its Appl., 1972, vol. 6, pp. 94-101, https://doi.org/10.1007/BF01077511

\bibitem{NgGreb13} B.-T. Nguyen,  D. S. Grebenkov, Localization of Laplacian Eigenfunctions in Circular, Spherical, and Elliptical Domains. //SIAM J. Appl. Math., 2013, vol. 73, No. 2, pp. 780-803, DOI: https://www.jstor.org/stable/23479950

\bibitem{Popov_G} G.S. Popov, Quasimodes for the Laplace Operator and Glancing Hypersurfaces,  // Microlocal Analysis and Nonlinear Waves, The IMA Volumes in Mathematics and its Applications,  Volume 30, Ed. Michael Beals, R. Melrose and J. Rauch, Springer Verlag, 1991

\bibitem{Popov20} M.M. Popov, A new concept of interference-type surface waves for smooth strictly convex surfaces embedded in three-dimensional space // Scientific seminar Notes of St.-Petersburg Math. Institute, 2020, vol. 493, p. 301-313 (In Russian)

\bibitem{Popov20-1} M.M. Popov, On the coordination of the integral asymptotics of surface waves of the interference type with the wave field of their source. // Scientific seminar Notes of St.-Petersburg Math. Institute, 2020, vol. 493, p. 314-322 (In Russian) 

\bibitem{MF} V.P. Maslov, M.V. Fedoriuk, Semi-Classical Approximation in Quantum Mechanics. // Springer, 2001,  Softcover reprint of the original 1st ed. 1981 edition

\bibitem{Peierls} R.E. Peierls, Quantum theory of Solids. // Clarendon Press, Oxford, 2001

\bibitem{Dobr83} S.Yu. Dobrokhotov, Maslov’s Methods in the Linearized Theory of Gravitational Waves on a Liquid Surface. // Sov. Phys. Dokl., 1983, vol. 28, pp. 229-231; Russian version: Dokl. Akad. Nauk SSSR, 1983, vol. 269, no. 1, pp. 76-80. 

\bibitem{GrDobrSergT16} V.V. Grushin, S.Yu. Dobrokhotov, S.A. Sergeev, B. Tirozzi. Asymptotic Theory of Linear Water Waves in a Domain with Nonuniform Bottom with Rapidly Oscillating Sections. //Russian Journal of Mathematical Physics, 2016, Vol. 23, N. 4, pp. 455-474, https://doi.org/10.1134/S1061920816040038

\bibitem{Serg22} S.A. Sergeev, Asymptotic Solution of the Cauchy Problem with Localized Initial Data for a Wave Equation with Small Dispersion Effects. //  Differential Equations, 2022, vol. 58, No. 10, pp. 1376-1395, DOI: 10.1134/S00122661220100081

\bibitem{BrGrDobr12} J. Br\"{u}ning, V.V. Grushin, S.Yu. Dobrokhotov, Averaging of linear operators, adiabatic approximation, and pseudodifferential operators. // Math. Notes., 2012, vol.  92, pp. 151-165, https://doi.org/10.1134/S0001434612070188

\bibitem{DobrMinShl18} S.Yu. Dobrokhotov, D.S. Minenkov, S.B. Shlosman, Asymptotics of Wave Functions of the Stationary Schr\"{o}dinger Equation in the Weyl Chamber. // Theor. Math. Phys., 2018, vol. 197, pp. 1626-1634, https://doi.org/10.1134/S0040577918110065

\bibitem{DobrMinNeiSh19} S.Yu. Dobrokhotov, D.S. Minenkov, A.I. Neishtadt, S.B. Shlosman, Classical and Quantum Dynamics of a Particle in a Narrow Angle, // Regul. Chaotic Dyn., 24:6 (2019),  704-716, https://doi.org/10.1134/S156035471906008X

\bibitem{Olver} F.W.J. Olver, D.W. Lozier, R.F. Boisvert, C.W. Clark, NIST Handbook of Mathematical Functions. // NIST, Cambridge University Press, 2010

\bibitem{Olver58} F.W.J. Olver, Uniform asymptotic expansions of solutions of linear second-order differential equations for large values of a parameter. // Phil. Trans. Roy. Soc. (London), Series A, 1958,  vol. 250, N. 984, pp. 479-517, https://doi.org/10.1098/rsta.1958.0005

\bibitem{Erdelyi} A. Erd\'{e}lyi, Asymptotic Solutions of Differential equations with Transition Points or Singularities. // Jour. of Math. Phys., 1960, vol. 1, N. 1, pp. 16-26, https://doi.org/10.1063/1.1703631

\bibitem{ADNTs}  A.Yu. Anikin, S.Yu. Dobrokhotov, V.E. Nazaikinskii, A.V. Tsvetkova. Uniform Asymptotic Solution in the Form of an Airy Function for Semiclassical Bound States in One-Dimensional and Radially Symmetric Problems. // Theor. Math. Phys., 2019, vol. 201, pp. 1742-1770, https://doi.org/10.1134/S0040577919120079

\bibitem{M} V.P. Maslov, Operational Methods. // Moscow., 1976.

\bibitem{DNS_UMS} S.Yu. Dobrokhotov,  V.E. Nazaikinskii, A.I. Shafarevich, Efficient asymptotics of solutions to the Cauchy problem with localized initial data for linear systems of differential and pseudodifferential equations. // Russian Mathematical Surveys, 2021, Vol. 76, issue 5, pp. 745-819, https://doi.org/10.1070/rm9973



\bibitem{ArnlKozlNeist} V.I. Arnold, V.V. Kozlov, A.I. Neishtadt, Mathematical Aspects of Classical and Celestial Mechanics, 3rd Ed. // Springer,  2006, 518 pages

\bibitem{StShN} V.E. Nazaikinskii, V.E. Shatalov, B.Yu. Sternin, Methods of Noncommutative Analysis. Theory and Applications. // Walter de Gruyter, Berlin, 1996.

\bibitem{Pressey} A. Pressey, Elementary Differential Geometry, Second Ed. // Springer, 2021

\end{thebibliography}
\end{document}